\documentclass[12pt,preprint]{aastex}
\slugcomment{Accepted to {\it AJ}, 2007 May 30}
\shortauthors{Looper}
\shorttitle{11 New Ts}

\received{2007 March 28}
\begin{document}

\title{Discovery of 11 New T Dwarfs in the Two Micron All-Sky Survey, 
Including a Possible L/T Transition Binary}

\author{Dagny L. Looper\footnote{Visiting Astronomer at the Infrared 
Telescope Facility, which is operated by the University of Hawaii under 
Cooperative Agreement no. NCC 5-538 with the National Aeronautics and 
Space Administration, Science Mission Directorate, Planetary Astronomy 
Program.}}
\affil{Institute for Astronomy, University of Hawai'i, 2680 Woodlawn Dr,
  Honolulu, HI 96822; dagny@ifa.hawaii.edu}

\author{J. Davy Kirkpatrick$^1$}
\affil{Infrared Processing and Analysis Center, MS 100-22, California 
    Institute of Technology, Pasadena, CA 91125}

\author{Adam J. Burgasser$^1$}
\affil{MIT Kavli Institute for Astrophysics \& Space Research, 77
    Massachusetts Ave, Building 37-664B, Cambridge, MA 02139}

\begin{abstract}

We present the discovery of 11 new T dwarfs, found during the course of
a photometric survey for mid-to-late T dwarfs in the 2MASS Point Source
Catalog and from a proper motion selected sample of ultracool dwarfs in 
the 2MASS Working Database.  Using the NASA Infrared Telescope
Facility SpeX spectrograph, we obtained low-resolution (R$\sim$150)
spectroscopy, allowing us to derive  
near-infrared spectral types of T2$-$T8.  One of these new T dwarfs, 2MASS
J13243559+6358284, was also discovered independently by Metchev et al.\, in prep. .
This object is spectroscopically peculiar and possibly a
binary and/or very young ($<$~300 Myr).  We specifically attempted to 
model the spectrum of this source as a composite binary to reproduce its 
peculiar spectral characteristics.  The latest-type object in our sample 
is a T8 dwarf, 2MASS J07290002$-$3954043, now one of the four latest-type T
dwarfs known.  All 11 T dwarfs are
nearby given their spectrophotometric distance estimates, 
with 1 T dwarf within 10 pc and 8 additional T dwarfs within
25 pc, if single.  These new additions increase the 25 pc census of T
dwarfs by $\sim$14\%.  Their proximity offers an excellent
opportunity to probe for companions at closer separations than are
possible for more distant T dwarfs.  

\end{abstract}


\section{Introduction}

The T spectral class \citep{2002ApJ...564..421B,2002ApJ...564..466G} 
is currently the coolest populated class on the MK
system, with effective temperatures ranging from $\sim$1400 K 
 at the L/T transition down to $\sim$700 K for 
the coolest known T dwarf \citep{2004AJ....127.3516G,2004AJ....127.2948V,2005ARA&A..43..195K}.  
The transition between the L and T spectral classes is
characterized by the onset of CH$_4$ absorption in the near-infrared.  
By mid-type T,
the spectral class is dominated by CH$_4$, H$_2$O, and collision induced
absorption by H$_2$ (CIA H$_2$) in the near-infrared 
\citep{1994ApJ...424..333S,1997A&A...324..185B, 2006ApJ...639.1120K}.  

Since the discovery of the T prototype Gl 229B
\citep{1995Natur.378..463N,1995Sci...270.1478O} in 1995, $\sim$100 T dwarfs have
been identified\footnote{See http://dwarfarchives.org for a full list of
T dwarfs.}.  These objects are our coolest neighbors, spanning the mass
and temperature gap between giant planets and the lowest mass stars.
Unlike extrasolar planets, we can directly image and often
spectroscopically observe T dwarfs in the field and as companions to
more massive stars.  Therefore, these objects offer invaluable testbeds
for extrasolar giant planet atmospheric models.

Due to their intrinsic faintness, the census of T dwarfs in the Solar
Neighborhood remains largely incomplete.  To date, two surveys have
provided the largest contributions to the currently known population:
the Sloan Digitized Sky Survey (SDSS, \citealt{2000AJ....120.1579Y}), 
a wide-field optical survey, and the Two Micron All Sky Survey (2MASS;
\citealt{2006AJ....131.1163S}), 
a near-infrared all sky survey.  T dwarfs emit most of their flux in the
near-infrared; however, due to the $J-K_s$ color
degeneracy of early-type T dwarfs and M dwarfs, recovery
of early-type T dwarfs by the 2MASS survey has proven difficult.  The SDSS
survey, while successful at identifying early-type (hotter) 
T dwarfs using far-optical color selections, is not as sensitive 
to late-type (cooler) T dwarfs, which emit very little flux at visual 
wavelengths.

To expand the census of the Sun's coolest neighbors, we have conducted a
photometric survey for mid-to-late type T dwarfs using the 2MASS Point Source
Catalog, searching both deeper than previous surveys for T dwarfs 
\citep{1999ApJ...522L..65B,2000ApJ...531L..57B,2000AJ....120.1100B,2002ApJ...564..421B,
2003AJ....125..850B,2003AJ....126.2487B,2004AJ....127.2856B,2005AJ....130.2326T,2005AJ....130.2347E} 
as well as in the Galactic Plane.  
We present the discovery of eight new T dwarfs from this search,
including the discovery of a very nearby (d$\sim$8.4
pc) T8 dwarf.  We also present the first three T dwarfs from our 2MASS
proper motion search, all of which are early-type T dwarfs.  We describe
these two searches in $\S$2.  In $\S$3, we describe 
the spectroscopic observations obtained with the
NASA Infrared Telescope Facility 3.0 m SpeX spectrograph and the
characterization of this new set of T dwarfs.  
In $\S$4, we discuss the current census of the Solar Neighborhood.  
Finally, in $\S$5 we give our conclusions. 

\section{Observations}

\subsection{Target Selection}

The eleven new T dwarfs we report here were discovered in two different
searches using Two Micron All Sky Survey (2MASS) data: 
(1) a 2MASS photometric search for mid-to-late type T dwarfs and 
(2) a 2MASS proper motion survey for ultracool dwarfs.  

\subsubsection{2MASS Photometric Selection of Mid-to-Late Type T Dwarfs}

This search was designed to identify mid-to-late type T dwarfs and to 
be complementary to previous searches by Burgasser et al.\ using the
2MASS database.  Their most recent search selected sources with 
$|$b$|\ge$~15$^\circ$, J~$\le$~16.0, J$-$H~$\le$~0.3 or H$-$K~$_s\le$~0, and no
optical counterpart within a 5$\arcsec$ radius in the USNO-A2.0 catalog
or by visual inspection of DSS I and II images.  We extended our search
half a magnitude deeper in J-band for the same area of sky
($|$b$|\ge$~15$^\circ$, 16~$<$~J~$\le$~16.5) and searched the galactic plane
($|$b$|<$~15$^\circ$), a portion of sky completely unexplored by
Burgasser et al.  We also modified our color selection criteria, described
below, to decrease the number of false positives followed-up
spectroscopically. 

Our search was broken into three parts defined by galactic latitude: 
(1) $|$b$|\ge$~15$^\circ$, (2) 10$^\circ$~$\le~|$b$|<$~15$^\circ$, 
and (3) $|$b$|<$~10$^\circ$.  All three searches had the following 
keywords (in parentheses) in common --

1. Not a cataloged minor planet at the time the 2MASS Point Source Catalog
   was constructed (mp\_flg=0);

2. No contamination by galaxies (gal\_contam=0);

3. No artifact contamination or source confusion (cc\_flg like '000');

4. No optical counterpart within a 5$\arcsec$ radius in the USNO-A2.0
   Catalog (nopt\_mchs=0) or by visual inspection of DSS II
   I-band images\footnote{Images acquired from 
   http://cadcwww.dao.nrc.ca/cadcbin/getdss.};

5. A non-null detection in both J- and H-bands (j\_cmsig
   is not null and h\_cmsig is not null);

6. Sky positions not coincident with the Large Magellanic Cloud, Small 
   Magellanic Cloud, or 47 Tuc.

The J-band magnitude and near-infrared color cuts for each of the three
galactic latitude searches are defined in Table 1.  
For (1) $|$b$|\ge$~15$^\circ$ and (2) 10$^\circ$~$\le~|$b$|<$~15$^\circ$, 
we used the same color selection ((J$-H\le$~0) or 
(J$-H\le$~0.3 and H$-K_s\le$~0)).  This color selection is 
similar but more restrictive than that used by \cite{2003AJ....125..850B} 
(J$-H\le$~0 or H$-K_s\le$~0) and was modified 
from that search because of the high incidence of M
dwarfs ($\sim$89\%) and low incidence of T dwarfs in spectroscopic follow-up
($\sim$11\%, \citealt{2004AJ....127.2856B}).  Burgasser et al$.$ had relaxed
color selection criteria to allow detection of early-type T dwarfs ($<$~T4).  
For (3) $|$b$|<$~10$^\circ$, we further restricted our color
selection (J$-H\le$~0) and brightness limit (J~$\le$~16) 
due to the high density of sources in the galactic plane and
searched only outside the galactic center, $|$l$|\ge$~20$^\circ$. These 
color selections effectively limited our search to SpT~$>$~T4 (see Figure 1).  

Of the three searches, the galactic plane search: 
(3) $|$b$|<$~10$^\circ$ and $|$l$|\ge$~20$^\circ$,
had the lowest rate of transients (40\%, asteroids, spurious
detections, flare events, etc$.$, 
see Table 2) and, after confirmation, highest rate of 
T dwarfs (100\%).  The three new T dwarfs from this search are 2MASS 
0602+40\footnote{We abbreviate all of our
discoveries from ``2MASS Jhhmmss[.ss]$\pm$ddmmss[.]s'' to ``2MASS hhmm$\pm$dd''
in the text for ease of reading but give their full designations in
Table 3.}, 2MASS 1007$-$45, and 2MASS 2154+59 (see Table 3).
In comparison, (1) $|$b$|\ge$~15$^\circ$ had a transient rate 
of $\sim$79\% and, after confirmation, T dwarf
rate of $\sim$58\%, with 4 new T dwarfs identified: 2MASS 0510$-$42, 
2MASS 1215$-$34, 2MASS 1615+13, and 2MASS 2154$-$10 (see Table 3).   
Three previously known T dwarfs were also recovered: SDSS 
1630+08 (T5.5, \citealt{2006AJ....131.2722C}), SDSS 
1758+46 (T6.5, \citealt{2004AJ....127.3553K}), and 
SDSS 2124+01 (T5, \citealt{2004AJ....127.3553K}, see Table 4).
The search (2) 10$^\circ$~$\le~|$b$|<$~15$^\circ$ had a 
transient rate of $\sim$67\% and, after confirmation, 
a T dwarf rate of 50\%, with 1 new T dwarf identified: 2MASS 0729$-$39
(see Table 3).  The lower transient and higher T dwarf rate 
for the galactic plane search arises
from the better photometry (J~$\le$~16) and more restrictive color
selection (J$-H\le$~0) used.  

The total incidence of transients in these three searches was $\sim$71\%.  In
Figure 2, we show a histogram of the fraction of transients versus
ecliptic latitude.  The vast majority (84\%) are located
within 30$^\circ$ of the ecliptic.  The remaining 16\% are
likely highly inclined asteroids or image artifacts.  We list all 50
transients in Table 2.

By contrast, the 8 newly identified and 3 confirmed 
T dwarfs are distributed nearly evenly
across ecliptic latitude ($\beta$; see Figure 2).  For all confirmed
candidates, we find a fraction of 0.45 for M dwarfs and 
0.55 for T dwarfs, 
an improvement of five-fold over Burgasser et al$.$'s search
despite more uncertain photometry for $|$b$|\ge$~10$^\circ$ and
16.0~$<$~J~$\le$~16.5.  This increased rate of T dwarf detection occurred 
because our color selection lies further from the
locus of M dwarfs, leading to less scattering from uncertain photometry
of faint M dwarfs into our color selection (see Figure 1).  
To date, we have completed follow-up of 67
of 70 candidates visible from
IRTF\footnote{IRTF: $\sim-$48$^\circ$~$<~\delta~\le$~+69$^\circ$
56$^\prime$ (hard limit).}
and have 11 candidates too northerly or southerly to be observed by IRTF.  
Of these remaining 14 candidates, 11 have $|\beta|>~30^\circ$.  Taking
into account the photometry and ecliptic latitudes of these remaining
sources, we estimate that 4$-$5 of these remaining candidates are T
dwarfs.  We have programs on IRTF/SpeX, Gemini-North/NIFS, and
SOAR/OSIRIS to obtain
imaging/spectroscopy of all remaining 14 candidates and will present
the results in a future paper.

\subsubsection{2MASS Proper Motion Survey}

We have carried out a proper motion survey for ultracool dwarfs using $\sim$4700
sq$.$ degrees of multi-epoch data in the 2MASS Working Database.  We give
a brief outline of the search criteria here and refer the reader to
Looper et al.\, in prep, for full details of this search.  For this dataset, we
selected all objects with a 
proper motion exceeding {0$\farcs$2} yr$^{-1}$, a time
difference between epochs of $\Delta$t $\ge$ 0.2 yr, a positional
difference exceeding 0$\farcs$4 between the first and last epochs, and  
J $\le$ 16.5.  We also required no optical counterpart within a 
5$\arcsec$ radius in
the USNO-A2.0 catalog \citep{1998USNO2.C......0M} or by visual inspection of 
Digitized Sky Survey (DSS) I and II V- or R-band images.  For
particularly bright objects, we required large R$-$J colors (R$-$J $>$ 6).
This search resulted in $\sim$140 candidates.  To date, we have
spectroscopically followed-up $\sim$120 of these objects, and we  
report on follow-up of three of these objects here: 2MASS 1106+27, 
2MASS 1324+63, and 2MASS 1404$-$31 (see Table 3).
2MASS 1324+63 was independently identified as
a T dwarf by Metchev et al.\, in prep.  Both 2MASS 1106+27 and 2MASS
1404$-$31 are new discoveries.  We also identified
two previously discovered T dwarfs: 2MASS 0939$-$24 (T8,
\citealt{2005AJ....130.2326T}) and SDSSp 1346$-$00 (T6.5,
\citealt{2000ApJ...531L..61T}, see Table 4).

\subsection{Imaging: UH 2.2m/ULBCAM \& IRTF 3.0m/SpeX}

We conducted follow-up imaging to cull the large number of transients
from our 2MASS photometric search sample, using the UH 2.2m\footnote{Located
on the summit of Mauna Kea, Hawai'i.}/ULBCAM and the IRTF 3.0m$^6$/SpeX
cameras.  Our 2MASS proper motion search sample did not
require follow-up imaging since second epoch 2MASS images confirmed the
existence of our targets.  We imaged a total of 67 candidates during
several runs described below.

The Ultra Low Background Camera (ULBCAM\footnote{See 
http://www.ifa.hawaii.edu/88inch/2.2-meter-technical.htm}) 
is a mosaic infrared camera with four
2048~$\times$~2048 arrays, a 17$^\prime~\times$~17$^\prime$ field-of-view
(FOV), and a 0$\farcs$25 pixel scale.  However, we used only the top right
array, with an individual FOV of 
8$^\prime$ 32$\arcsec~\times$~8$^\prime$ 32$\arcsec$, because it has the 
highest quantum efficiency and lowest number of bad pixels
of the four arrays.  We imaged candidates with ULBCAM on 
09$-$12 \& 16 Feb 2006 UT and 24 Jul 2006 UT with clear and photometric
conditions on all nights and good seeing ($\lesssim$~1$\arcsec$ at J-band).  
We used the J-band filter for all observations and dithered between 
exposures, where individual exposure times varied from 20 to 180 s.
All images were first bias subtracted and then pair-wise subtracted to
give the final image.  All resultant images were deeper than the 2MASS
J-band image for the same field.  

SpeX has a 512~$\times$~512 array InSb camera with a
60$\arcsec~\times$~60$\arcsec$ FOV and a 0$\farcs$12 pixel
scale \citep{2003PASP..115..362R}.  We used this camera to
image candidates during 31 May 2006 UT, 17$-$18 Aug 2006 UT, 01$-$02 Sep 2006
UT, 17 Nov 2006 UT, and 08$-$09 \& 20$-$21 Dec 2006 UT.  Conditions for the
31 May 2006 UT run were clear with 1$\arcsec$ seeing at J-band.  Conditions
were also clear with 0$\farcs$6$-$0$\farcs$7 seeing at J-band for the 17$-$18 Aug
2006 UT and 01$-$02 Sep 2006 UT runs.  Conditions for the 17 Nov 2006 run
ranged from clear to light cirrus with 1$\farcs$0 seeing at K-band and
similar conditions for the 08$-$09 \& 20$-$21 Dec 2006 UT runs with
0$\farcs$5$-$0$\farcs$8 seeing at K-band.  Using the J-band filter, we took 60
sec exposures with a single offset for pair-wise subtraction.  In all
cases where neighboring stars fell in the camera's FOV, we were able to
verify that the resultant images were deeper than the 2MASS J-band image
for the same field.  Candidates absent in our two imaging campaigns are
listed in Table 2.

\subsection{Spectroscopy: IRTF 3.0m/SpeX}

Spectroscopic observations were obtained using the IRTF/SpeX
spectrograph in low-resolution mode, covering
$\sim$0.7$-$2.5 $\mu$m in a single order on the 1024$\times$1024 InSb
array.  We used the 0$\farcs$5 slit, resulting in R$\sim$150, and
rotated the slit to the parallactic angle to minimize slit losses.  For
accurate sky subtraction, we nodded along the slit in ABBA cycles.  
Exposure times
for science targets ranged from 120$-$180 s to maximize signal-to-noise
and to minimize temporal variations of OH airglow, which can leave
a forest of residual lines for long exposures (see Figures 1, 2, \& 3
of \citealt{1992MNRAS.259..751R}).  

We obtained spectroscopic observations for a total of 20 science targets, 2 T dwarf standards, 
and 20 calibrator A0 V stars
between 08 Apr 2006 UT and 21 Dec 2006 UT.  We
list a log of our observations in Table 3.  All nights were clear or had
light cirrus.  All calibrator
A0 stars were observed either immediately before or after the science
target with a differential airmass less than 0.10 for accurate
flux calibration.  After observing the calibrator stars, we immediately
took internal flat-field and argon arc lamps for instrumental
calibration.  We employed standard reduction techniques using the
Spextool package version 3.2 \citep{2004PASP..116..362C,2003PASP..115..389V}.

\section{Analysis}

\subsection{M Dwarfs}

Near-infrared spectroscopy of candidates confirmed via second-epoch
imaging (see $\S$2.2) reveals nine of them as M dwarfs.  To classify these
spectra, we compared their overall morphology, particularly the strength
of their H$_2$O absorptions at 1.30$-$1.51 and 1.75$-$2.05 $\mu$m, to a set of
known M dwarfs: Gl 229A (M1), Gl 411 (M2), Gl 388 (M3), Gl 213 (M4), Gl
51 (M5), Gl 406 (M6), vB8 (M7), vB10 (M8), LHS 2924 (M9), and BRI
0021$-$0214 (M9.5) \citep{2005ApJ...623.1115C}.
We list the spectral types (M3$-$M7) of
these nine M dwarfs in Table 3 and estimate our accuracy as $\pm$1
subtype, except for those types denoted with a colon due to spectra
with poor signal-to-noise.  None of these 9 M dwarf spectra differ
markedly from our comparison M dwarfs, suggesting that the unusual
near-infrared colors of these M dwarfs shown in Figure 1 are due to
large photometric errors instead of spectroscopic peculiarities.  

\subsection{T Dwarfs}

\subsubsection{Classification \& Kinematics}

In Figure 3, we show finder charts for all 11 new T dwarfs constructed
from 2MASS All-Sky Release
Survey\footnote{See http://irsa.ipac.caltech.edu/Missions/2mass.html.} 
J-band images.  We classified these T dwarfs by comparing their overall spectral
morphologies to near-infrared T dwarf primary standards (see Figure 4) 
defined by \cite{2006ApJ...637.1067B}:  
SDSS J120747.17+024424.8 (T0 std), SDSSp
J083717.22$-$000018.3 (T1 std), SDSSp
J125453.90$-$012247.4 (T2 std), 2MASS
J12095613$-$1004008 (T3 std), 2MASSI
J2254188+312349 (T4 std), 2MASS
J15031961+2525196 (T5 std), SDSSp
J162414.37+002915.6 (T6 std), 2MASSI
J0727182+171001 (T7 std), and 2MASSI 
J0415195$-$093506 (T8 std).
These were also observed with IRTF/SpeX
using the 0$\farcs$5 slit, so they have identical resolution
(R$\sim$150) to our spectra (\citealt{2006ApJ...637.1067B}, and this paper).
We calculated near-infrared H$_2$O and CH$_4$ spectral indices defined
in \cite{2006ApJ...637.1067B} for the 11 new T dwarfs (see Table 5).  The
indirect spectral types (calculated by comparing each index measurement
to that of each standard and finding the nearest
subtype or half subtype match)  
are in excellent agreement ($\le$~0.5 subtypes) with our
direct spectral classification except for 2MASS 0729$-$39, which has two
indices that are classified as $>$~T8 and are not included in the mean
indirect spectral type. 

We find a spectral type range of T2 to
T8 and a wide range in color of $-$0.6~$\lesssim$~J$-$K$_s\lesssim$~1.5.  
We were able to identify early-type T dwarfs in 2MASS because we did not rely 
on near-infrared color selections in our proper motion search.
With the exception of 2MASS 1324+63 (T2:pec) and 2MASS
0729$-$39 (T8pec), these T dwarfs show CH$_4$ and H$_2$O absorption
strengths throughout the near-infrared regime consistent with
standards of similar spectral types (see Figure 4).  In Figure 5, we show the
near-infrared (J, H, K$_s$) colors along with optical (i, z) colors, when
available, of these 11 T dwarfs in comparison to all known T dwarfs with
individual magnitude errors of less than 0.3 mag.  

The near-infrared colors of these new T dwarfs span a broad range, but
generally track with the trend in color as a function of spectral type.
While 2MASS 1324+63 lies along the redder edge of the
dispersal relative to objects of similar spectral type 
and 2MASS 0729-39 lies along the bluer edge relative to objects of
similar spectral type, they are both
within one sigma of their own photometric errors from other sources in
the dispersion.  

Although the bottom panel of far optical minus near-infrared
colors has very few sources, 2MASS 1324+63 is indistinct from the
scatter of other T2 dwarfs.  On the other hand, 
2MASS 1106+27 is the only T2.5 dwarf with a measured i-band
color in SDSS but is distinctly ($>1\sigma $) bluer than all T2 dwarfs 
and is the bluest
T2.5 dwarf in z$-$J.  Although 2MASS 1106+27 is the second brightest T
dwarf in SDSS (z=17.70$\pm$0.03, see Table 4), it was not found in
previous searches of SDSS for T dwarfs because the photometry for a
nearby faint source was registered instead.  To get the SDSS colors for
this object, we had to use the Navigate Visual Tool\footnote{See 
http://cas.sdss.org/astrodr5/en/tools/chart/navi.asp.} to find 2MASS
1106+27.  We used the Explore radio button and recorded the parameters
of this object, which we fed to an SQL query, requesting
the psf magnitudes (see Table 4).  The i and z magnitudes of 2MASS
1324+63 and 2MASS 1615+13, we obtained from the SDSS Catalog in a
standard fashion, requesting the 'psf\_mags' and 'psf\_magerrs' (see Table
4). 

Spectrophotometric distance estimates were calculated using
the spectral types derived above and the \cite{2006ApJ...647.1393L} spectral type
vs.\ magnitude relation (excluding known binaries) and are given along
with other spectrophotometric properties in Table 4.  These distances
range from $\sim$8.4$-$29.0 pc.  For
the three early-type T dwarfs from the 2MASS proper motion survey, we were
able to calculate proper motions and position angles from the multi-epoch
data (see Table 6 for kinematic properties).  We were also able to do
the same for 2MASS 1615+13, which was detected in z-band in the SDSS
Catalog.  The proper motions along with the distance
estimates yielded tangential velocity estimates of
$\sim$36$-$66 km s$^{-1}$ for these four objects, which 
are near or exceeding the median
V$_{tan}$~=~39.0 km s$^{-1}$ for T dwarfs found by \cite{2004AJ....127.2948V}.  The
tangential velocities of these four T dwarfs are nearly twice that of
the median tangential velocities of L dwarfs (V$_{tan}$~=~24.5 km
s$^{-1}$, \citealt{2004AJ....127.2948V}).  Our proper motion sample is biased
towards these higher velocity objects; i.e., at a distance of 20 pc, we
would only detect objects with 
V$_{tan}~\gtrsim~4.74~\times~0\farcs4$ yr$^{-1}~\times$~20 pc~=~38 km
s$^{-1}$.  We use $0\farcs4$ yr$^{-1}$ in this estimate since $0\farcs4$
is the motion floor for the survey and the mean epoch difference is near
1 yr.

\subsection{Spectroscopically Peculiar Sources}

Two of the T dwarfs identified in this study have spectral properties
that are somewhat unusual compared to the spectral standards.  We
discuss these peculiar sources in detail.

\subsubsection{2MASS 1324+63 (T2:pec)} 

While the spectrum of 2MASS 1324+63 has an overall spectral morphology
similar to the T2 standard, particularly from the red optical to the
blue slope of the H-band peak (1.5 $\mu$m), at longer wavelengths it is
considerbaly more red, with weaker CH$_4$ absorption at 1.6 $\mu$m and a
brighter K-band peak (2.1 $\mu$m).  This motivates our classification of
this source as peculiar.  There are a handful of peculiar T dwarfs
known, some of which are known to be binary (e.g., 2MASS
J05185995$-$2828372; 
\citealt{2004ApJ...604L..61C}) or suspected to be affected by gravity effects
(e.g., 2MASSI J0937347+293142;
\citealt{2002ApJ...564..421B,2006ApJ...639.1095B,2004AJ....127.3553K}).  We consider both of
these possibilities for 2MASS 1324+63.

To examine the binary hypothesis, we constructed a suite of synthetic
binaries from L8--T8 spectral templates with equivalent SpeX prism
spectra\footnote{Synthetic spectra available upon request.}.  
Our technique is similar to that used in
\cite{2005ApJ...634L.177B,2006ApJS..166..585B} and \cite{2006ApJ...639.1114R}.
Our spectral templates include the primary near-infrared T dwarf
spectral standards defined by \cite{2006ApJ...637.1067B} 
and the L dwarfs 2MASSW J1632291+190441 (L8, \citealt{1999ApJ...519..802K}) 
and 2MASSW J0310599+164816 (L9, \citealt{2000AJ....120..447K}).  We
augmented the SpeX spectra from \cite{2006ApJ...637.1067B} with new spectra
for SDSS J120747.17+024424.8 (T0; \citealt{2002AJ....123.3409H}) and SDSSp
J083717.22$-$000018.3 (T1; \citealt{2000ApJ...536L..35L}).  
Most of these sources are
unresolved in high angular resolution imaging
(\citealt{2006ApJS..166..585B,2006ApJ...637.1067B} and references therein), with
the exception of 2MASS 0310+16, which is a near equal-brightness binary
\citep{2005prpl.conf.8571S}.  The L9, T0, T3, and T7 templates have yet to be
imaged at high angular resolution.

Integrated light spectra for artificial binaries were constructed by
first absolutely flux calibrating the A+B spectral components.
We computed  
a correction factor to scale the spectrum of each component such that it
had the appropriate absolute J-band magnitude given by its spectral
type.  The correction factor is given by,

C$_J=10^{-0.4\,M_J} \frac{\int \lambda\,
f_\lambda^{Vega}(\lambda)\,T_J(\lambda)\,d\lambda}{\int \lambda\,f_\lambda^{obs}(\lambda)\,T_J(\lambda)\,d\lambda}$

where both the J-band Vega 
spectrum\footnote{Downloadable from Sandy Leggett's site $-$ 
ftp://ftp.jach.hawaii.edu/pub/ukirt/skl/filters/vega.obs.}, 
$f_\lambda^{Vega}$, and the 2MASS J-band transmission profile\footnote{See 
http://spider.ipac.caltech.edu/staff/waw/2mass/opt\_cal/jrsr.tbl.html.}, 
$T_J(\lambda)$, 
have been interpolated onto the wavescale of each observed 
component.  The absolute J-band magnitude, M$_J$, was computed from
the \cite{2006ApJ...647.1393L} spectral type vs absolute magnitude relation,
excluding known binaries.  We calculated and multiplied this correction factor
for both A+B components.  Then we interpolated the wavelength scale of
the B component onto the wavelength scale of the A component and added
the two fluxes for the combined synthetic spectrum.  For quantitative
reference, we calculated H$_2$O and CH$_4$ spectral indices    
for the entire suite (see col$.$ 2$-$6, Table 7) but classify the
synthetic spectra on overall morphology in comparison with spectral 
standards (see col$.$ 9 -- Direct, Table 7).  The spectral type of each 
synthetic combination computed by the indices is shown in col$.$ 10 of
Table 7 (Ind 1).  These indices are shown graphically in Figure 6 for 
the NIR synthetic spectra, NIR T dwarf standards, and the 11 new T
dwarfs.  

The synthetic binary spectra were compared to the spectrum of 2MASS
1324+63 visually.  The four closest matches are shown in Figure 7.
The best two matches are L9+T2 [T0.5] and L8+T5 [T2:].
The L9+T2 synthetic combination, while
providing a fairly good match throughout most of H- and K$_s$-bands and
having comparable H$_2$O absorption strength 
from 1.75$-$2.05 $\mu$m, has weaker H$_2$O absorption
from 1.3$-$1.5 $\mu$m, a weaker Y-band ($\sim$0.93$-$1.15 $\mu$m; 
\citealt{2002PASP..114..708H}) slope, and weaker CH$_4$ and
H$_2$O absorption from $\sim$1.1$-$1.2 $\mu$m.  The L8+T5 synthetic
combination, conversely, provides a good match in Y- and J-bands but has
overly strong absorption in H- and K-bands, which are primarily shaped
by CH$_4$ absorption and CIA H$_2$.  While none of these matches are
ideal, it should be noted that similar analysis for the resolved binary
DENIS-P J225210.73$-$173013.4 also fails to provide a perfect fit, even
though the source is known to be a binary \citep{2006ApJ...639.1114R}.  So
unresolved multiplicity may nevertheless play a significant role.

Another possibility is that 2MASS 1324+63 is a young T dwarf 
($<$~300 Myr), implying a lower mass, larger radius, and hence lower
surface gravity (low pressure photosphere).
One of the primary spectral shapers in K-band and, to a
lesser extent, H-band, is CIA H$_2$.  In a low pressure atmosphere, the
contribution of CIA H$_2$ would decrease and the spectrum would become
redder (J$-$K would increase).  This effect has been noted for
young L dwarfs (\citealt{2006ApJ...639.1120K}, Kirkpatrick et al.\, in prep).  
2MASS 1324+63 could also have a dustier photosphere than typical T2 dwarfs,
possibly caused by a lower gravity environment retarding the
precipitation of dust.  To date, the youngest spectroscopically
confirmed T dwarf\footnote{A candidate very low mass T dwarf in the $\sim$1 Myr
Orion cluster, S Ori 70, has been proposed by \cite{2002ApJ...578..536Z} and 
\cite{2003ApJ...593L.113M}.  However, this source remains controversial 
\citep{2004ApJ...604..827B}.} is 
HN Peg B \citep{2007ApJ...654..570L}, a T2.5 companion to HN Peg A (a G0 V star 
estimated to be $\sim$300 Myr old).  After $\sim$300 Myr, brown dwarf 
radii contract very little, differing by only $\sim$20\% in radius 
from their much older 3 Gyr counterparts \citep{1997ApJ...491..856B}.  While
this relatively young object is less massive than a 3 Gyr T2.5 dwarf, 
HN Peg B shows no spectroscopic deviations from 
a typical field T2.5 dwarf (See Figure 8 and \citealt{2007ApJ...654..570L}).  

2MASS 1324+63 is located in a region of
sky completely unoccupied by any currently known young moving group. 
It could be a member of an unidentified young moving group or it
could be older ($>$~100 Myr) than most moving groups (either a cast-off or 
from a
dispersed group).  Its tangential velocity (V$_{tan}$~=~45 km
s$^{-1}$) seemingly contradicts this, suggesting 
that it is of comparable age to older field T dwarfs.

\subsubsection{2MASS 0729$-$39 (T8pec)} 

This object is one of four spectroscopically classified 
T8 dwarfs, the latest spectral type for T
dwarfs currently known.  While the peculiarities of 2MASS
0729$-$39 are not as evident as those for 2MASS 1324+63, they are
certainly noticeable and worth further exploration here.   
This source is peculiar because it has some excess flux in the
Y-band peak while the J-band peak, the H$_2$O and CH$_4$
absorption from $\sim$1.1$-$1.2 $\mu$m, and the H$_2$O absorption from
1.3$-$1.5 $\mu$m matches well to the T8 standard.  There is also reduced 
flux in the H-band and K-band peaks compared to the T8 standard.  

These peculiarities are indicative of a higher pressure photosphere,
increasing K I wing absorption (causing slightly increased Y-band flux) 
and CIA H$_2$ absorption (causing depressed H- and K-band flux).  Such
features were previously noted for the high surface gravity and 
metal poor dwarf 2MASS 0937+29 (See Figure 2 of \citealt{2006ApJ...639.1095B}).  
2MASS 0729$-$39 may be a similarly relatively old and/or slightly
metal-poor T dwarf.

\subsection{Temperatures \& Gravities}

To examine the properties of our latest-type T dwarfs in further
detail, we made use of the semi-empirical spectral index technique of
\cite{2006ApJ...639.1095B} to estimate T$_{eff}$ and
$\log{g}$.  In brief, this method involves the comparison of
H$_2$O and color spectral ratios measured
on the spectrum of a late-type T dwarf
to the same ratios measured on theoretical condensate-free spectral models
from \cite{2001ApJ...556..357A} and \cite{2006ApJ...640.1063B}.
The latter are calibrated to
reproduce the measured indices for the near infrared SpeX prism
spectrum of Gliese~570D \citep{2000ApJ...531L..57B}, which has
parameters of T$_{eff}$ = 782--821~K,
log g = 4.95--5.23 and [Fe/H] = 0.09$\pm$0.04, based on
empirical measurements and evolutionary models 
\citep{2001ApJ...556..373G,2006ApJ...647..552S}.
The H$_2$O and color ratios are separately
sensitive to T$_{eff}$ and log g (for a given metallicity),
and thus break the degeneracy between these parameters,
which can then be used to infer mass and age with evolutionary models.
As condensate-free
spectral models generally provide poor fits for T$_{eff}$ $\gtrsim$ 1200~K, our
analysis is generally confined to sources $\gtrsim$T5
\citep{2004AJ....127.3516G}.  See \cite{2007astro.ph..1793B} and
\cite{2007ApJ...655..522L} for examples of this spectral technique in
application.

We were able to derive constraints for T$_{eff}$ and log g for five of the new
T dwarfs in our sample using the H$_2$O-J and K/H ratios (the latter defined
in \citealt{2006ApJ...639.1095B}).  Figure 9 illustrates the
fits for these sources, and Table 8 lists their estimated T$_{eff}$ and
log g ranges assuming an uncertainty of 10\% in the spectral ratio
measurement.  Note that additional systematic uncertainties of order
50~K and 0.2~dex should be included when interpreting these values.
T$_{eff}$ values are consistent with trends from parallax samples
\citep{2004AJ....127.3516G,2004AJ....127.2948V}, and show the expected
decline in T$_{eff}$ with later spectral type.  Note that the peculiar
T8, 2MASS 0729$-$39 has a relatively large surface gravity, consistent
with our interpretation of this source in $\S$~3.5.2.  2MASS 1007$-$45,
on the other hand, has a relatively low surface gravity, which is
consistent with its brighter K-band flux peak as seen in Figure~3
relative to the T5 standard.  Subsolar and supersolar metallicities
can also reproduce these effects, respectively
\citep{2006ApJ...639.1095B,2007astro.ph..1793B,2007ApJ...660.1507L,2007ApJ...656.1136S}. 

Using the solar metallicity evolutionary models of \cite{2001RvMP...73..719B}, 
we estimated masses and ages for these five sources; ranges
(without including systematic uncertainties) are given in Table 8.
As expected, low (high) surface gravities result in low (high) mass
and age estimates.  Since metallicity effects can mimic surface
gravity effects on the K-band (in both cases modulating the relative
opacity of H$_2$), care should be taken in interpreting these values.

\section{Discussion - The Solar Neighborhood}

To place our discoveries in context and to review the current state of
high-resolution imaging for the nearest T dwarfs, we have constructed
an up-to-date census of T dwarfs in the Solar Neighborhood ($<$~25 pc).
This list was constructed from the L \& T dwarf compendium maintained by
Kirkpatrick, Gelino, \& Burgasser\footnote{See http://dwarfarchives.org.}.
We computed spectrophotometric distance estimates for the entire list of
T dwarfs, using the M$_J$ vs near-infrared (NIR) SpT relation derived 
by \cite{2006ApJ...647.1393L} (excluding known binaries).  Those T dwarfs with 
trigonometric
parallaxes have their distance estimates superseded by these
measurements.  For T dwarfs known to be binaries but without parallaxes,
we have placed lower limits on their spectrophotometric distance
estimate.  Within 25 pc, this list includes a total of 72 T dwarfs,
including 9 new additions reported here (see Table 9).  These new
additions represent an $\sim$14\% increase in the 25 pc census of
T dwarfs.  Only 1 T dwarf
from the compendium meeting these criteria was excluded from this list
$-$ 2MASS J11263991$-$5003550 (NIR T0, \citealt{2007astro.ph..3808F}), which
Burgasser et al.\, in prep, demonstrate has a mid-type L dwarf optical spectral
morphology.  

Within 10 pc, there are currently
14 known T dwarfs, one of which, 2MASS 0729$-$39 (the 8th closest T
dwarf to the sun at 8.4 pc), is reported here.  Of these 14 T dwarfs, 5
are companions to more massive stars.  To date, over half of these 
T dwarfs have been followed-up with high-resolution imaging, with one 
revealed as a tight binary - the nearest T dwarf, $\epsilon$ Indi Bab
\citep{2003A&A...398L..29S, 2004A&A...413.1029M}.  This yields a binary fraction of
$\sim$13\% and a lower limit to the observed 
space density of 3$~\times~10^{-3}$ T dwarfs per pc$^{3}$ for the 10 pc
sample.  Metchev et al.\, in prep, performed Monte Carlo simulations,
predicting $\sim$30 T0--T8 dwarfs within 10 pc (a space density of
7$^{+3.2}_{-3.0}~\times~10^{-3}$ pc$^{-3}$), 
roughly twice the known population.  The inputs for these simulations
were based on a critical examination of T dwarf
candidates chosen from a cross-correlation of SDSS and 2MASS, which helps
to eliminate the 2MASS selection bias for early-T dwarfs and the SDSS
selection bias for very late-T dwarfs.
This density estimate
exceeds both that derived for T5--T8 dwarfs
(4.2$~\times~10^{-3}$ pc$^{-3}$; \citealt{2002ApJ...564..421B}) and for 
L dwarfs (3.8$~\times~10^{-3}$ pc$^{-3}$; \citealt{2007AJ....133..439C}).

Between 10 and 25 pc, there are currently 59 known T dwarfs, 8 of which
we report here.  One
caveat to this set, is that the distances we assigned to non-parallax
sources are biased into this sample if they are binary (i.e., they are
really further away, potentially outside 25 pc).  
The state of high-resolution imaging for this sample is
not nearly as complete as for d~$<$~10 pc, with only 23 out of the 59
known T dwarfs having follow-up.  Of these
23 T dwarfs, 7 have been identified as binaries (\citealt{2006ApJS..166..585B}
and references there-in), yielding a binary fraction of $\sim$30\% for
this sample.  This entire list
represents an interesting distance-limited subset of all known T dwarfs,
complete follow-up (both high resolution imaging and parallax
measurements) of which could greatly bolster binarity statistics
and lead to a more complete understanding of brown dwarf formation.
Based on the number of known T dwarfs within 10 pc, there should be 
$\gtrsim$~200 T dwarfs within 10~$<$~d~$<$~25 pc (Metchev et al.\, in prep, 
predict $\sim$430), so the current census is highly incomplete. 

We also note that, like the prototypes of the L and T spectral classes
(GD 165B, \citealt{1988Natur.336..656B} and Gl 229B, 
\citealt{1995Natur.378..463N,1995Sci...270.1478O}, respectively), which
are both companions, the prototype of the Y spectral class (hypothetical
dwarfs cooler than type T, \citealt{1999ApJ...519..802K}) could first be
identified as a companion to one of these T
dwarfs.  This should be ample motivation for the brown dwarf community
to continue high angular resolution imaging of our nearest cool neighbors.

\section{Conclusions}

We have reported on the discovery of 11 new T dwarfs found during the
course of two surveys using the 2MASS database.  Using low-resolution
prism spectroscopy (R$\sim$150) on IRTF/SpeX, we have classified these T
dwarfs in the near-infrared from T2$-$T8.  All of these T dwarfs are
nearby, with spectrophotometric distance estimates of $\sim$8.4$-$29.0
pc.  One of these T dwarfs, 2MASS J13243559+6358284, we type as T2:pec
and postulate that its spectroscopic peculiarities are the result of
reduced collision induced absorption by H$_2$, indicative of it being a
young object ($<$~300 Myr), or that it is an L/T transition binary.  
To provide models for suspected or known binaries, we constructed a
suite of synthetic spectra and computed their H$_2$O and CH$_4$ indices.
We have also reviewed the state of high resolution follow-up of T dwarfs in
the Solar Neighborhood with d~$<$~25 pc and make the case for completing
follow-up of this sample to improve the poorly sampled binary statistics
throughout the T spectral class for this distance-limited set.

\section{Acknowledgments}

DLL thanks J$.$ Rayner for advising her for part of this project and 
for a careful read of the manuscript. 
We would like to thank J$.$ Kartaltepe and Y$.$ Kakazu for imaging
several of our targets with the UH 2.2m/ULBCAM and for teaching DLL how
to use the instrument.  We thank S$.$ Metchev for helping us with the SDSS
Catalog, Mike Cushing for useful discussions, and Kevin Luhman for
kindly providing the spectrum of HN Peg B.  
We would also like to thank our telescope operators on IRTF:
D$.$ Griep \& B$.$ Golisch.  This paper uses data from the IRTF Spectral Library 
(\url{http://irtfweb.ifa.hawaii.edu/$\sim$spex/spexlibrary/IRTFlibrary.html})
and from \url{http://DwarfArchives.org}. 
This publication also makes use of data products from the Two Micron All Sky 
Survey (2MASS), which is a joint project of the University of
Massachusetts and the Infrared 
Processing and Analysis Center/California Institute of Technology, 
funded by the National Aeronautics and Space Administration and the 
National Science Foundation. DLL was a guest user of the 
Canadian Astronomy Data Centre, 
which is operated by the Herzberg Institute of Astrophysics, National 
Research Council of Canada.  This research has also made use of the 
NASA/IPAC Infrared Science Archive (IRSA), which is operated by the 
Jet Propulsion Laboratory, California Institute of Technology, under 
contract with the National Aeronautics and Space Administration. 
As all 
spectroscopic and imaging follow-up data were obtained from the summit 
of Mauna Kea, the authors wish to recognize and acknowledge the very 
significant cultural role and reverence that this mountaintop has always 
had within the indigenous Hawaiian community. We are most fortunate to 
have the opportunity to conduct observations on the summit.

%\noindent 

\begin{figure}
\epsscale{1.00}
\plotone{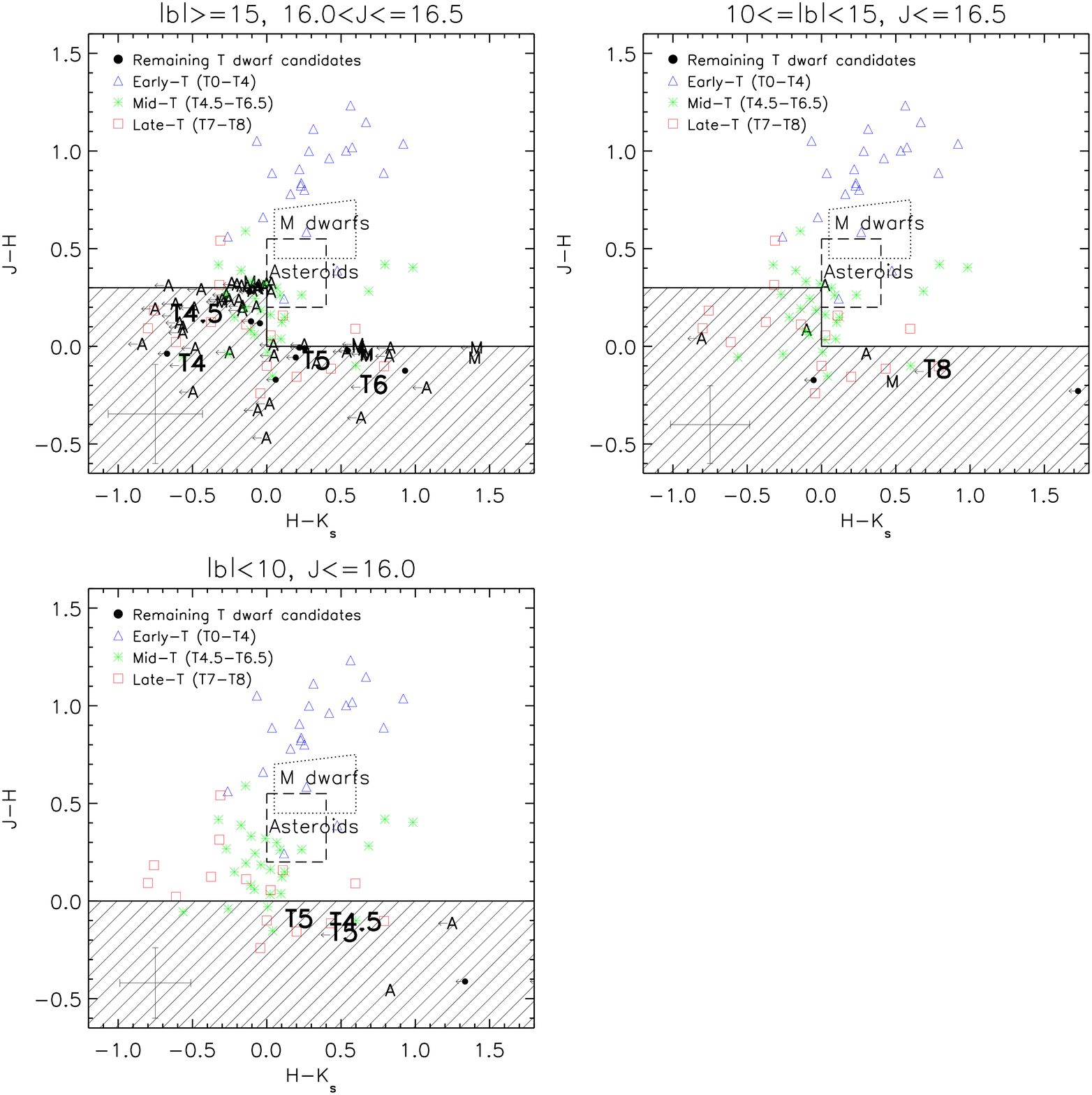}
\caption{Near-infrared color-color plots for the 2MASS photometric
search for mid-to-late type T dwarfs broken down by galactic latitude.  
Color selections are shown by the cross-hatched area.  
Known T dwarfs are plotted for reference, with early-type T dwarfs 
represented by blue triangles, mid-type T dwarfs by green asterisks, 
and late-type T dwarfs by red squares.  For
reference, the loci of M dwarfs and typical asteroids are outlined
(\citealt{2000AJ....120..447K,2004ApJ...605L.141G}, respectively).  Candidates
absent in follow-up imaging are denoted by ``A'' and spectroscopically
confirmed M dwarfs are denoted by ``M.''  All candidates needing follow-up
imaging/spectroscopy are denoted by solid circles and newly identified T
dwarfs are denoted by their spectral type.  For candidates with null
K$_s$ errors, arrows indicate an upper limit on their H$-K_s$ color.
Median error bars for all candidates with non-null color errors are shown in
the lower left of each plot.  
\label{fig1}}
\end{figure}

\begin{figure}
\epsscale{0.8}
\plotone{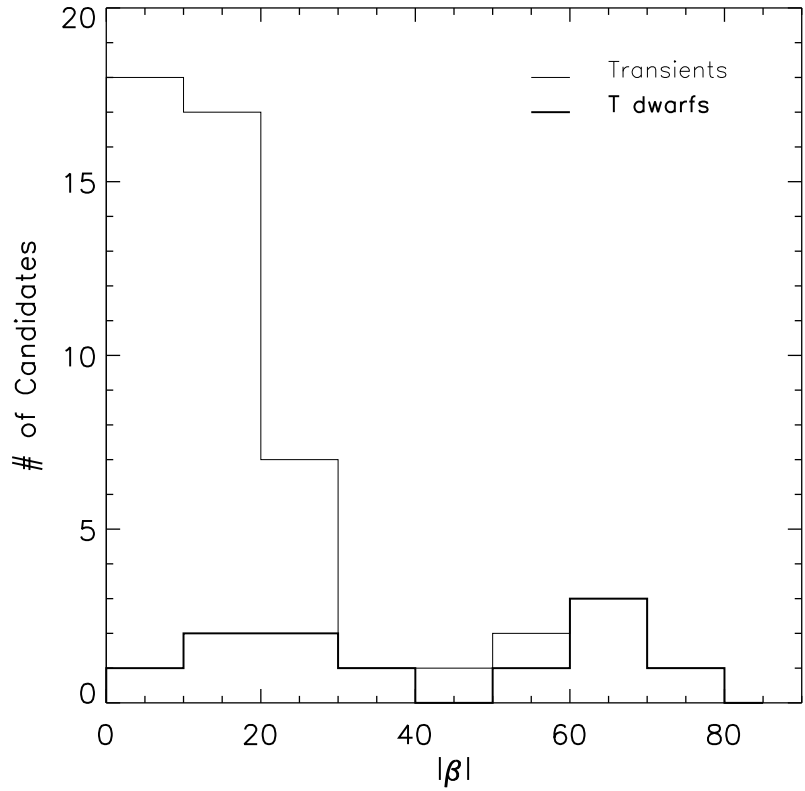}
\caption{A histogram for all 50 transients (light line) and all 8 new T
dwarfs and 3 recovered T dwarfs (heavy line), identified in the 2MASS 
photometric search for mid-to-late type 
T dwarfs, as a function of ecliptic latitude, $|\beta|$.  Note that,
while the T dwarfs are distributed nearly evenly across ecliptic
latitutudes, the vast majority of transients are concentrated near the
ecliptic, at $|\beta|\le$~30$^\circ$.
\label{fig2}}
\end{figure}

\begin{figure}
\epsscale{0.8}
\plotone{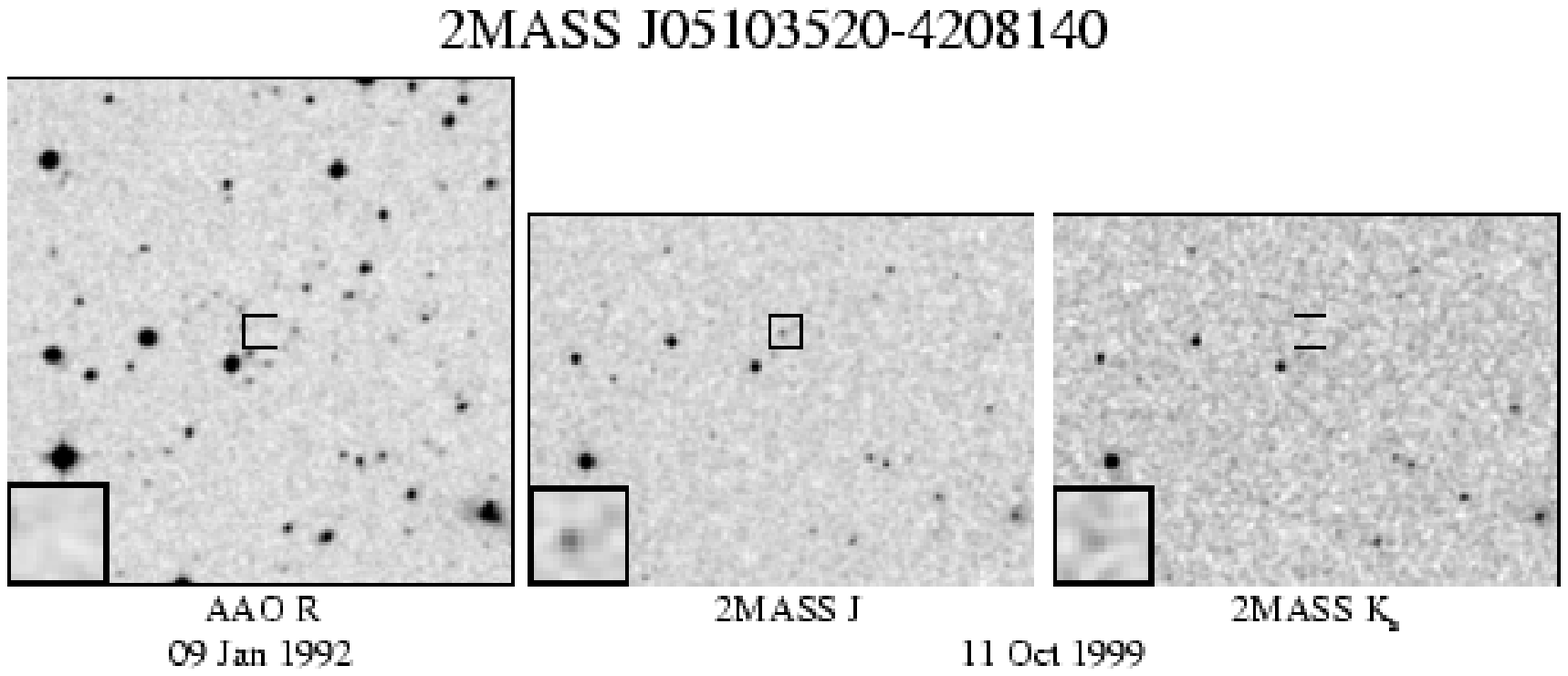}
\end{figure}

\begin{figure}
\epsscale{0.8}
\plotone{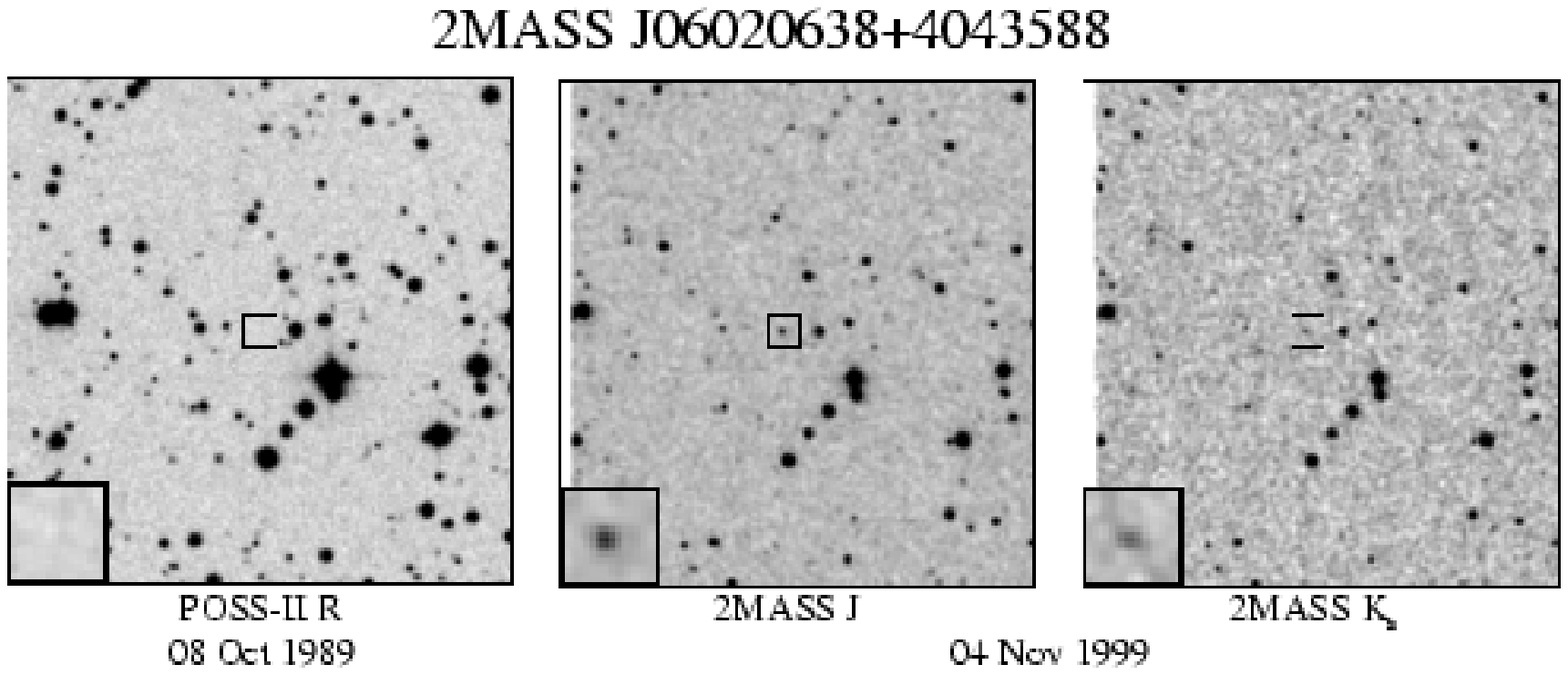}
\end{figure}

\begin{figure}
\epsscale{0.8}
\plotone{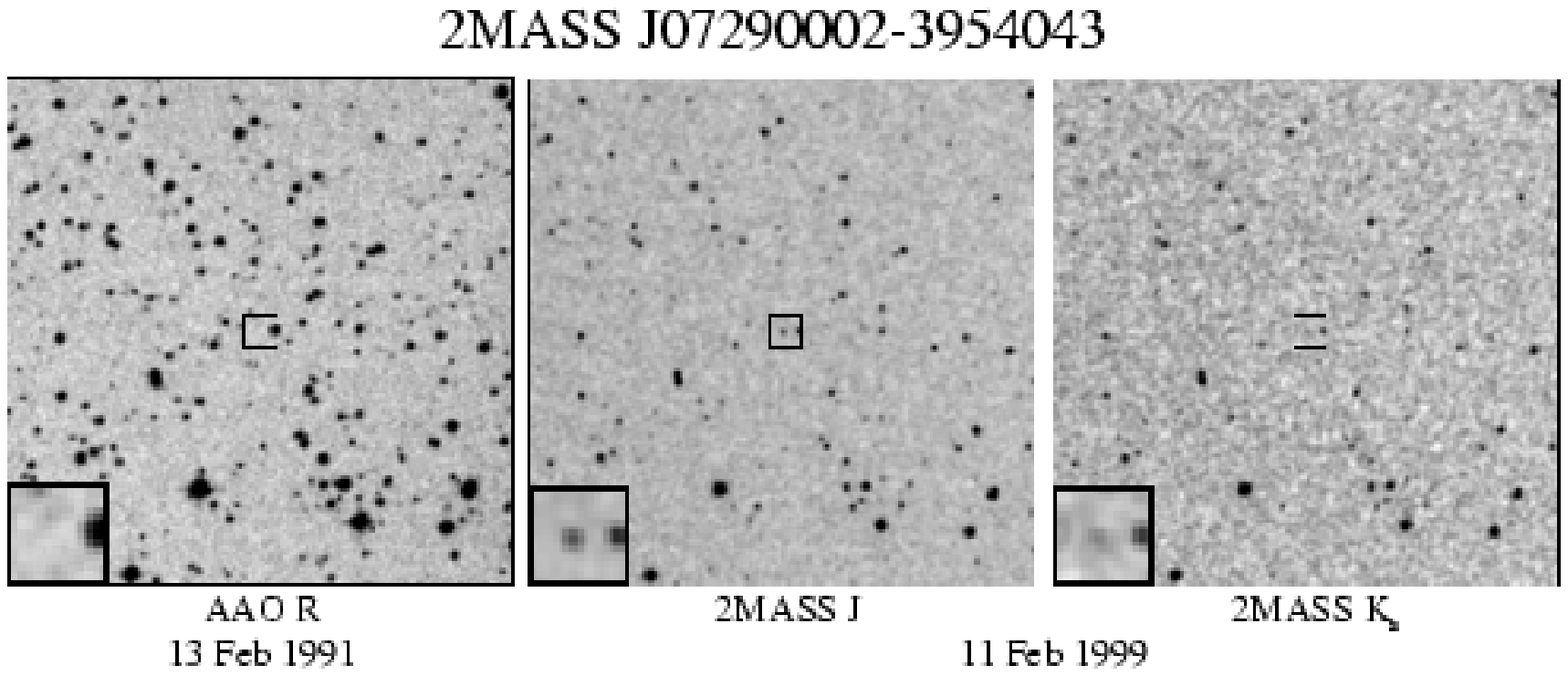}
\end{figure}

\begin{figure}
\epsscale{0.8}
\plotone{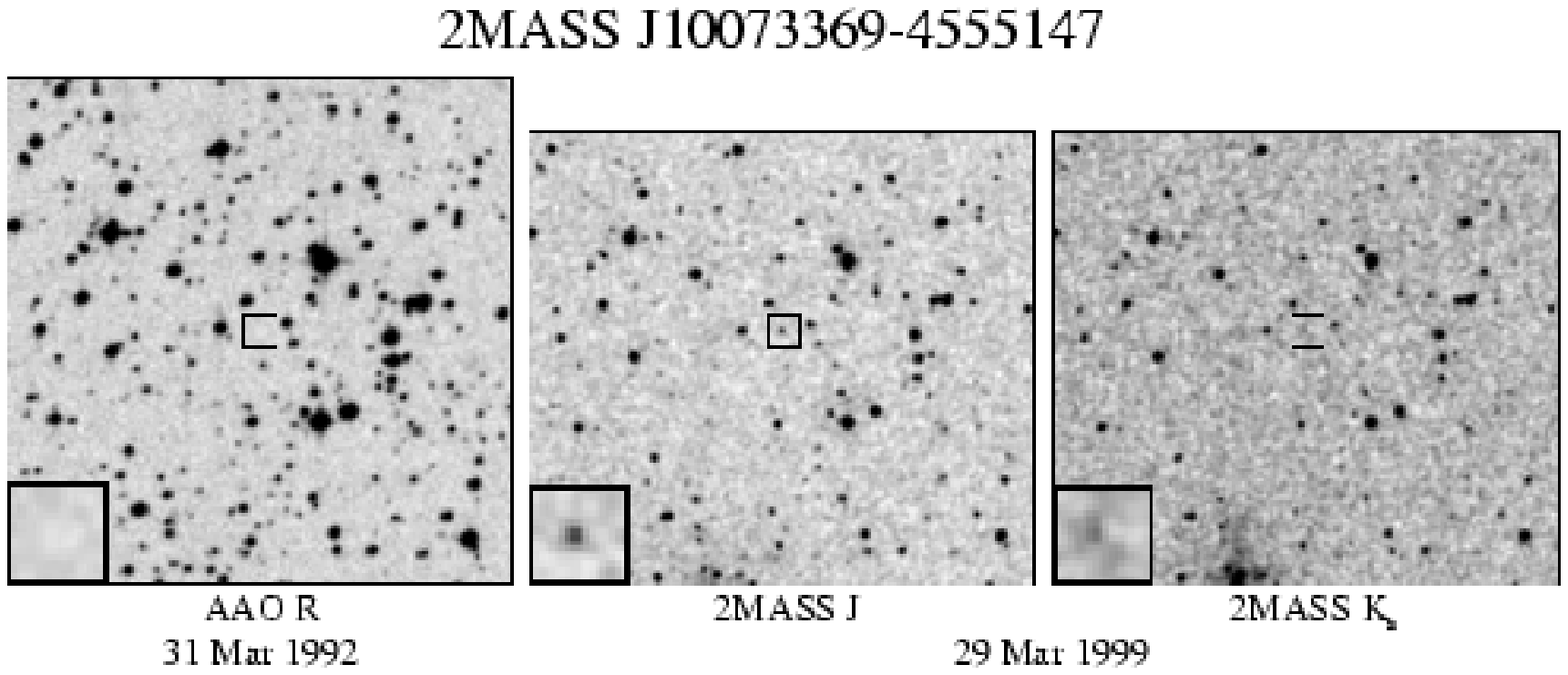}
\end{figure}

\begin{figure}
\epsscale{0.8}
\plotone{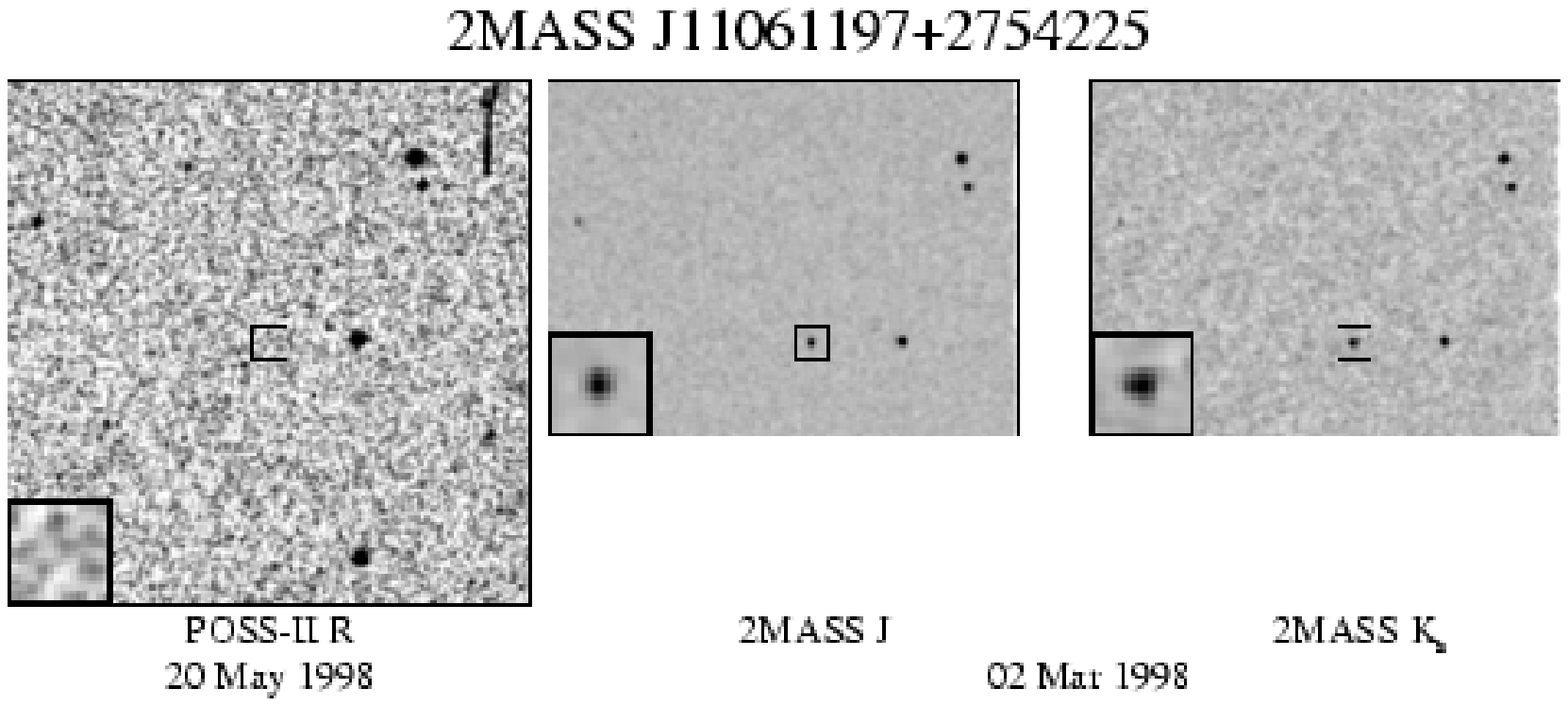}
\end{figure}

\begin{figure}
\epsscale{0.8}
\plotone{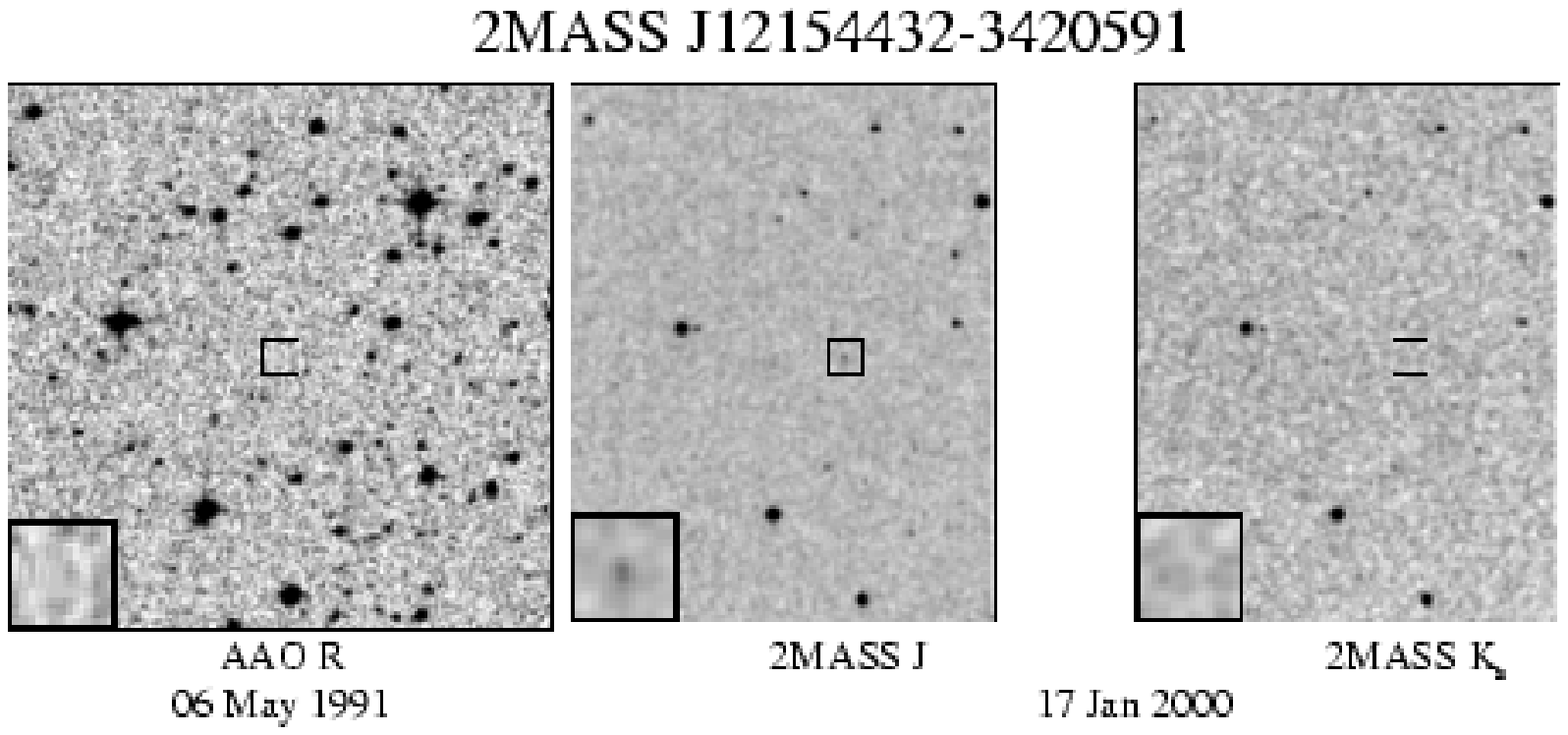}
\end{figure}

\begin{figure}
\epsscale{0.8}
\plotone{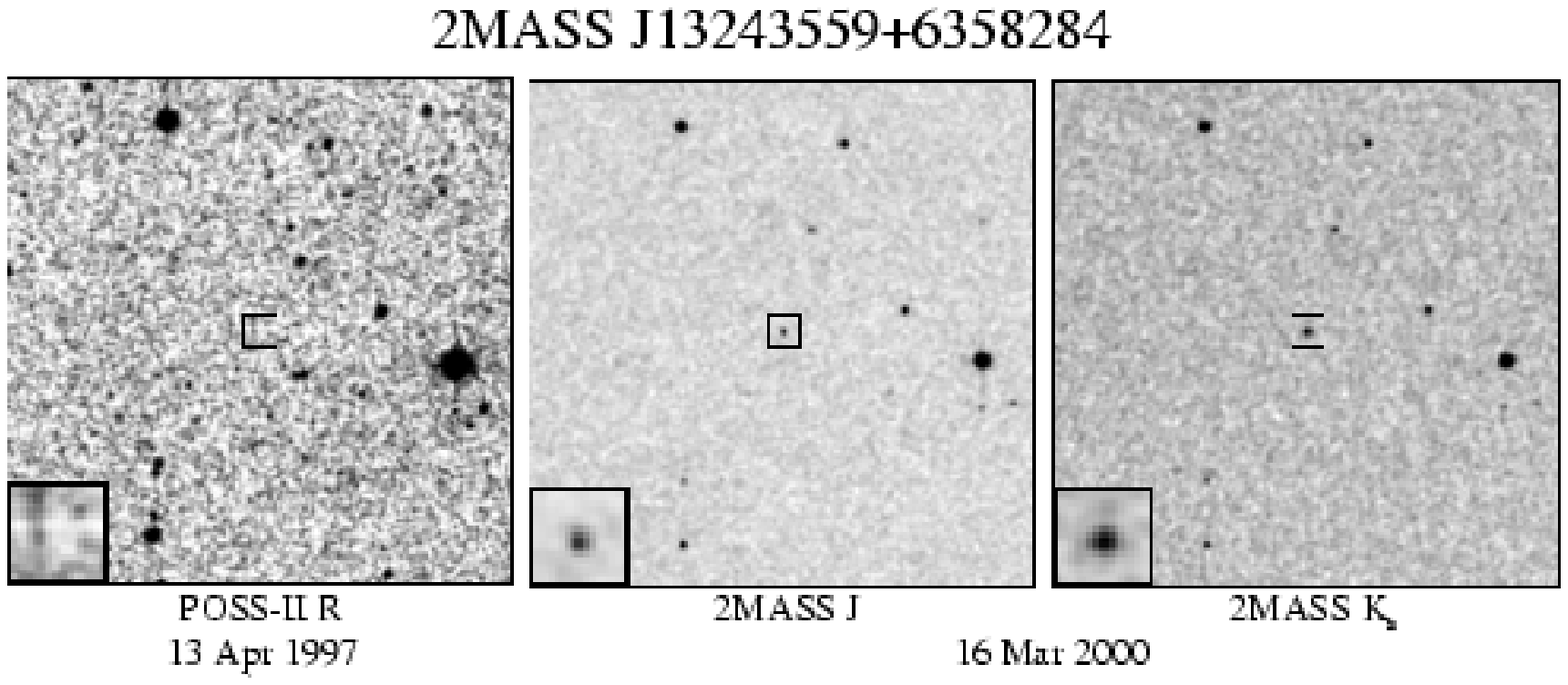}
\end{figure}

\begin{figure}
\epsscale{0.8}
\plotone{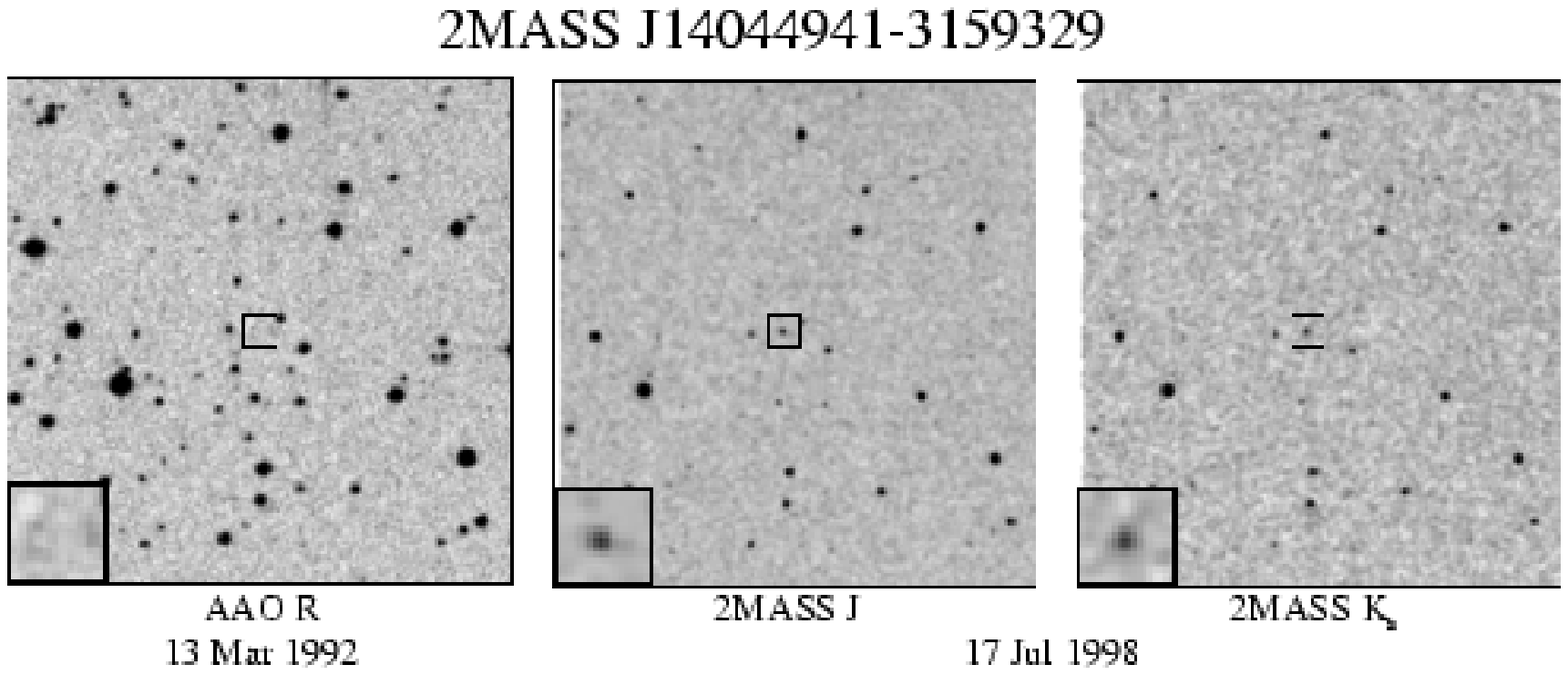}
\end{figure}

\begin{figure}
\epsscale{0.8}
\plotone{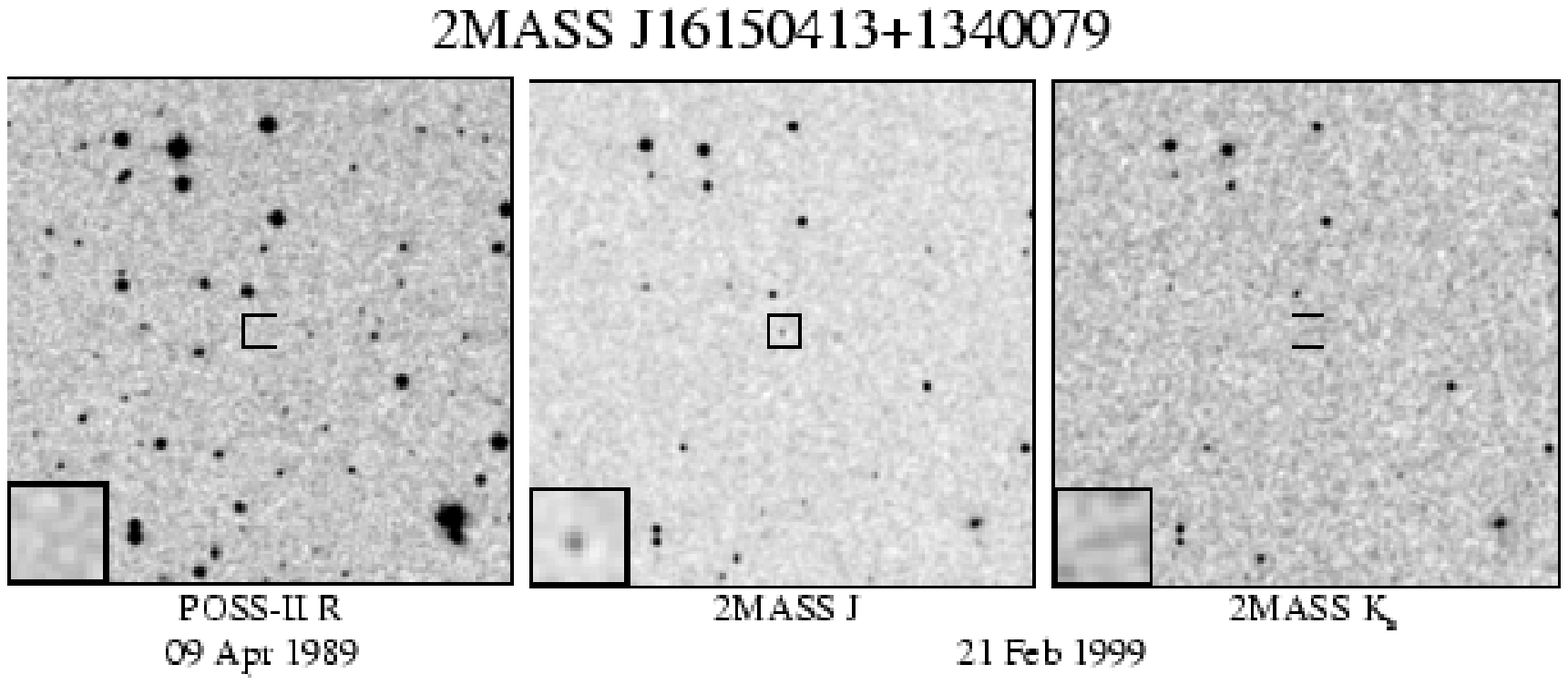}
\end{figure}

\begin{figure}
\epsscale{0.8}
\plotone{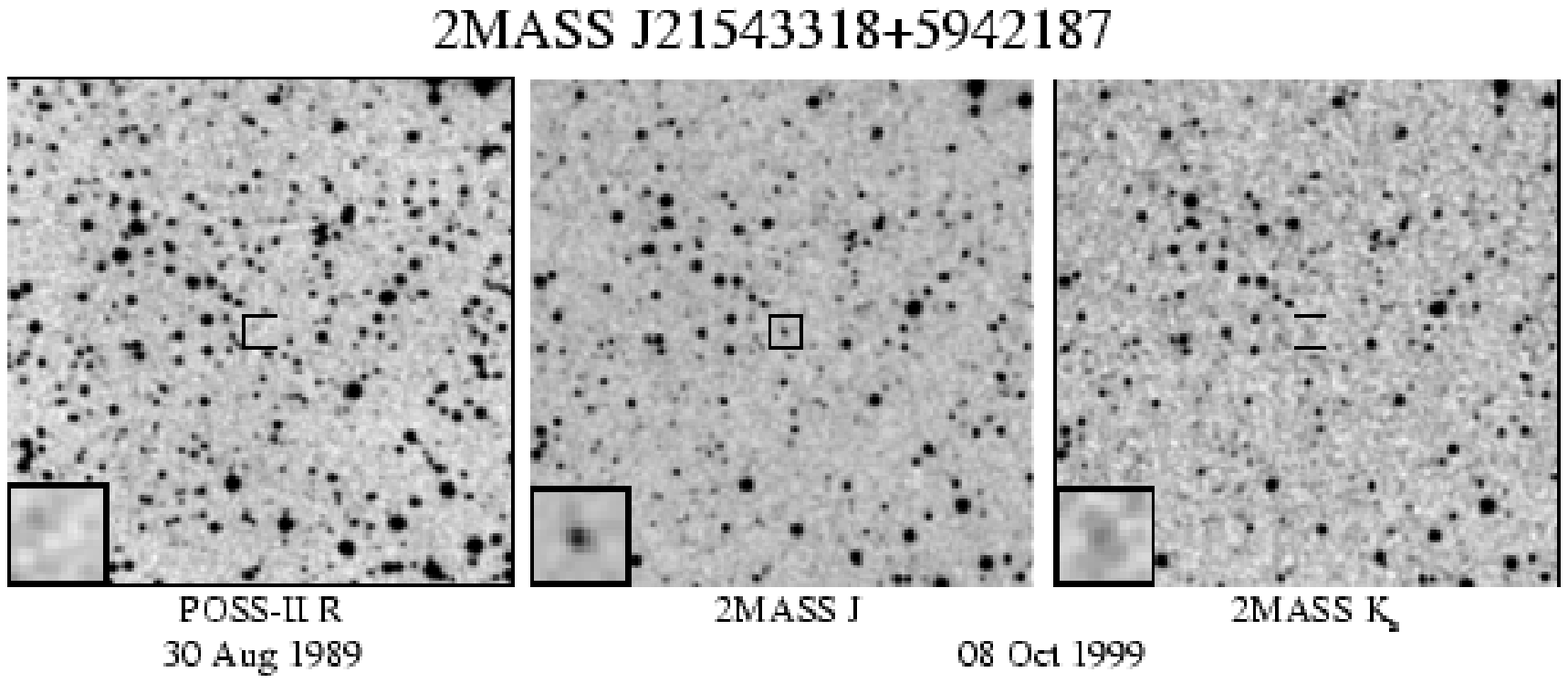}
\end{figure}

\begin{figure}
\epsscale{0.8}
\plotone{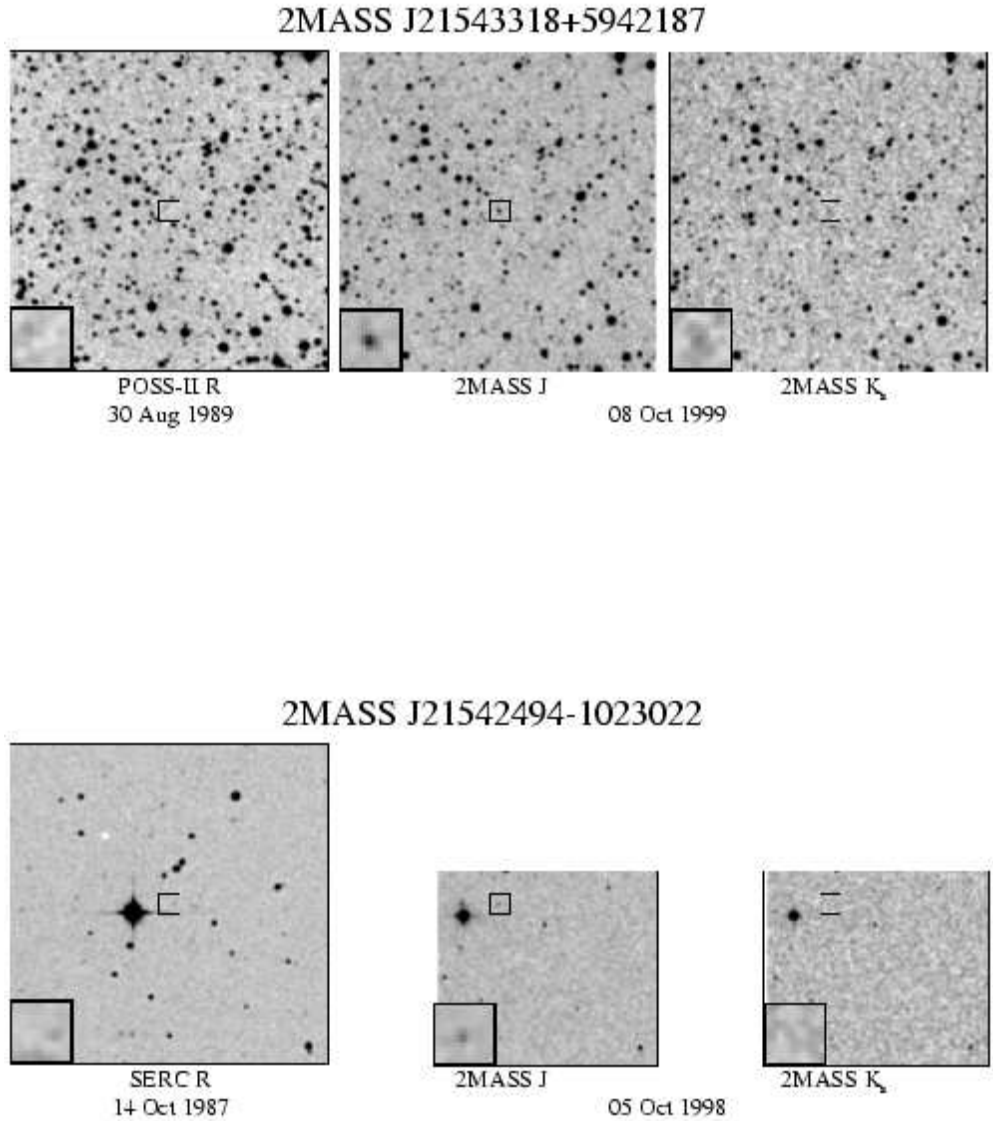}
\caption{Finder charts for the eleven new T dwarfs constructed from
DSS II R-band, 2MASS J-band, and 2MASS K$_s$-band images.  
Each image is nominally 300$\arcsec$ on a
side, although some images near the edge of a scan are smaller in size.
Inserts shown on the lower left are 10$\arcsec$ on a side.  
Epochs (UT) for each image are indicated along with the J2000
coordinates for each object represented by hhmm$\pm$ddmm.
\label{fig3}}
\end{figure}

\begin{figure}
\epsscale{1.0}
\plotone{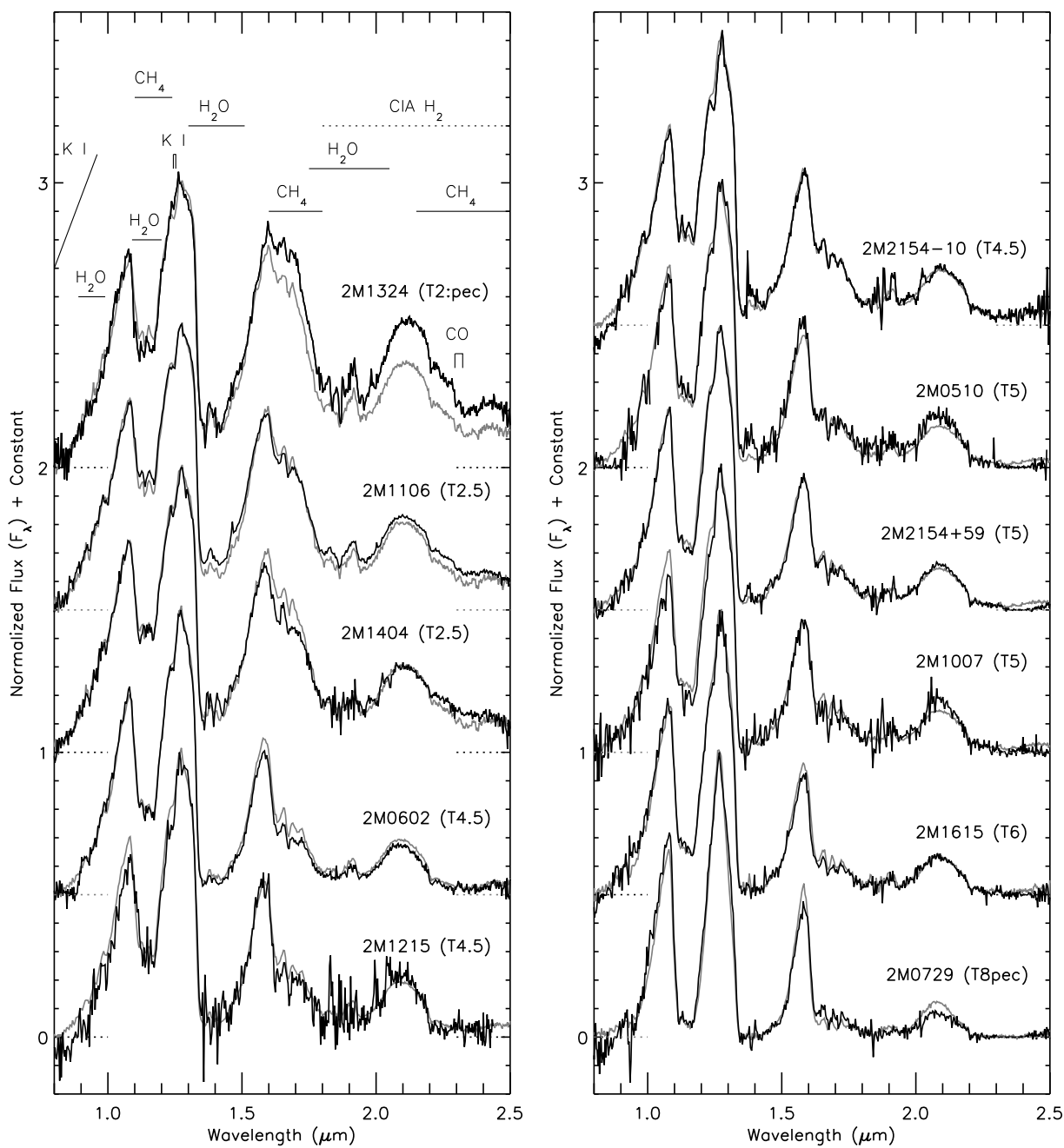}
\caption{IRTF/SpeX prism (R$\sim$150) spectra of the eleven newly
discovered T dwarfs (thick black lines) overplotted on their nearest
near-infrared spectral type match (thin gray lines).  Integer spectral
types are the near-infrared primary T dwarf standards defined by
\cite{2006ApJ...637.1067B}.  Half subtypes are from the
synthetic spectral library (see $\S$3.2).   
All spectra have been normalized at 1.27 $\mu$m
using a 0.01 $\mu$m window and are offset by half-integer values (dotted
lines show zero levels) for clarity.  We have indicated the key atomic
and molecular features.
\label{fig4}}
\end{figure}

\begin{figure}
\epsscale{1.0}
\plotone{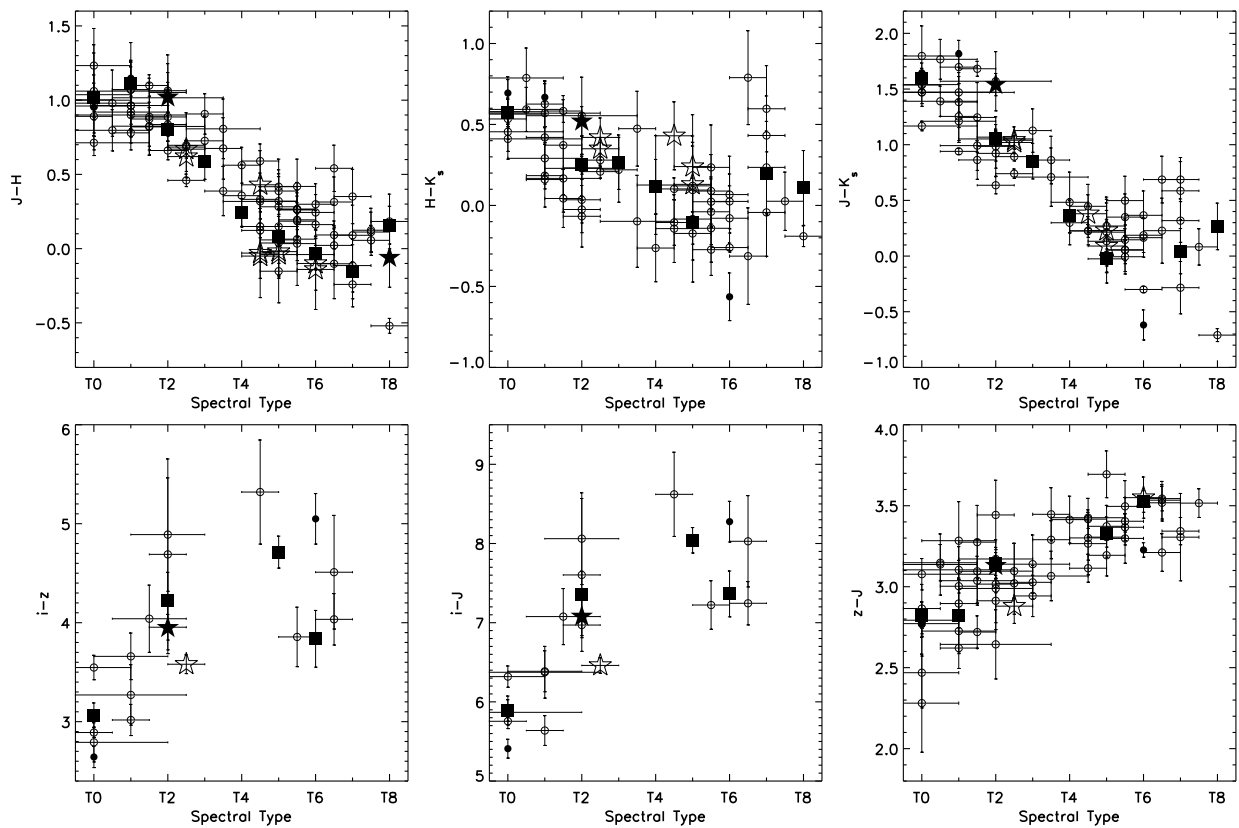}
\caption{Far optical and near-infrared colors of known T dwarfs (open
  circles), known peculiar T dwarfs (filled circles), T dwarf standards
  (filled squares), new T dwarfs (open stars), and new peculiar T dwarfs
  (filled stars) versus near-infrared spectral type.  
  All known T dwarfs shown have magnitude errors of
  less than 0.3 mag.  The i and z magnitudes are from the SDSS (AB) 
  Catalog, and the J, H, and K$_s$ magnitudes are from the 2MASS Catalog.
\label{fig5}}
\end{figure}

\begin{figure}
\epsscale{1.0}
\plotone{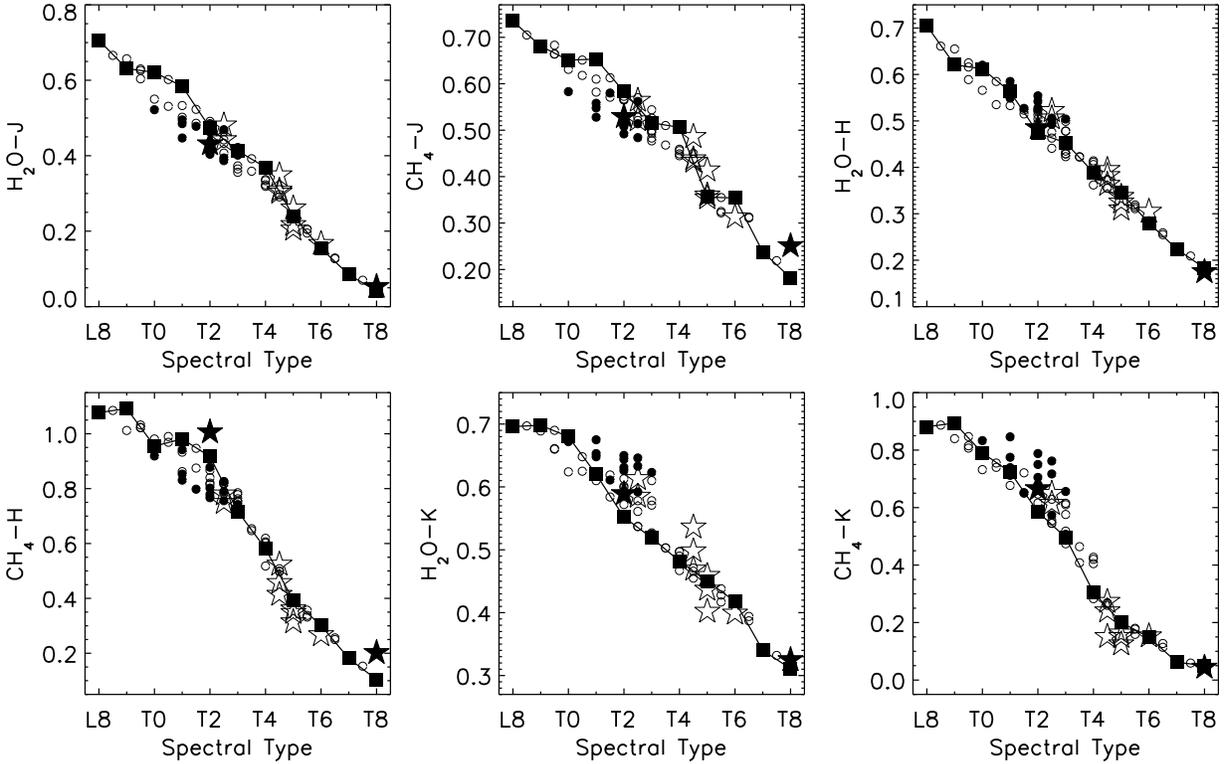}
\caption{Spectral index values of H$_2$O and CH$_4$ features versus
  near-infrared spectral type (determined by direct comparison) 
  as defined by \cite{2006ApJ...637.1067B}.  Solid squares represent the
  late-type L and T dwarfs we used to create the synthetic spectra, with
  a solid line connecting them.  
  Synthetic spectra are represented by circles, with peculiar or
  uncertain types denoted by filled circles.  The 11 new T dwarfs
  reported here are indicated by five-pointed stars, with peculiar
  sources denoted by filled stars.
\label{fig6}}
\end{figure}

\clearpage

\begin{figure}
\epsscale{0.6}
\plotone{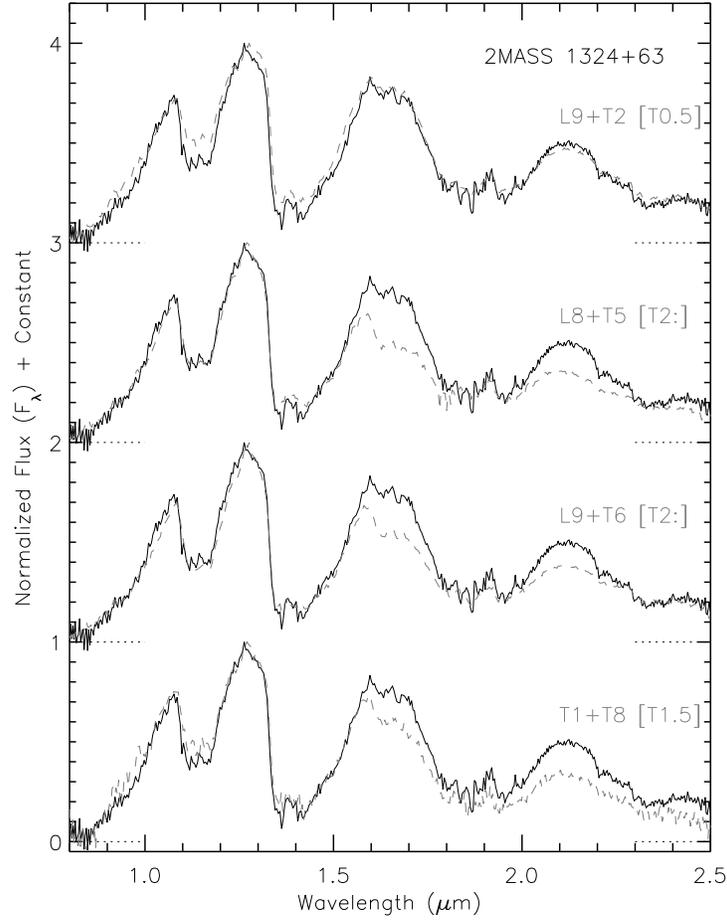}
\caption{Spectral template fitting of 2MASS 1324+63 (black)
overplotted with best synthetic spectral fits (gray) determined visually.  
The A+B components are labeled
and the integrated light spectral type (determined by direct comparison)
is shown in brackets.  All spectra
have been normalized at 1.27 $\mu$m and are offset by integer units
(zero levels are marked by dotted lines) for clarity.  Of these four,
the best fit is L9+T2 [T0.5].
\label{fig7}}
\end{figure}

\clearpage

\begin{figure}
\epsscale{1.0}
\plotone{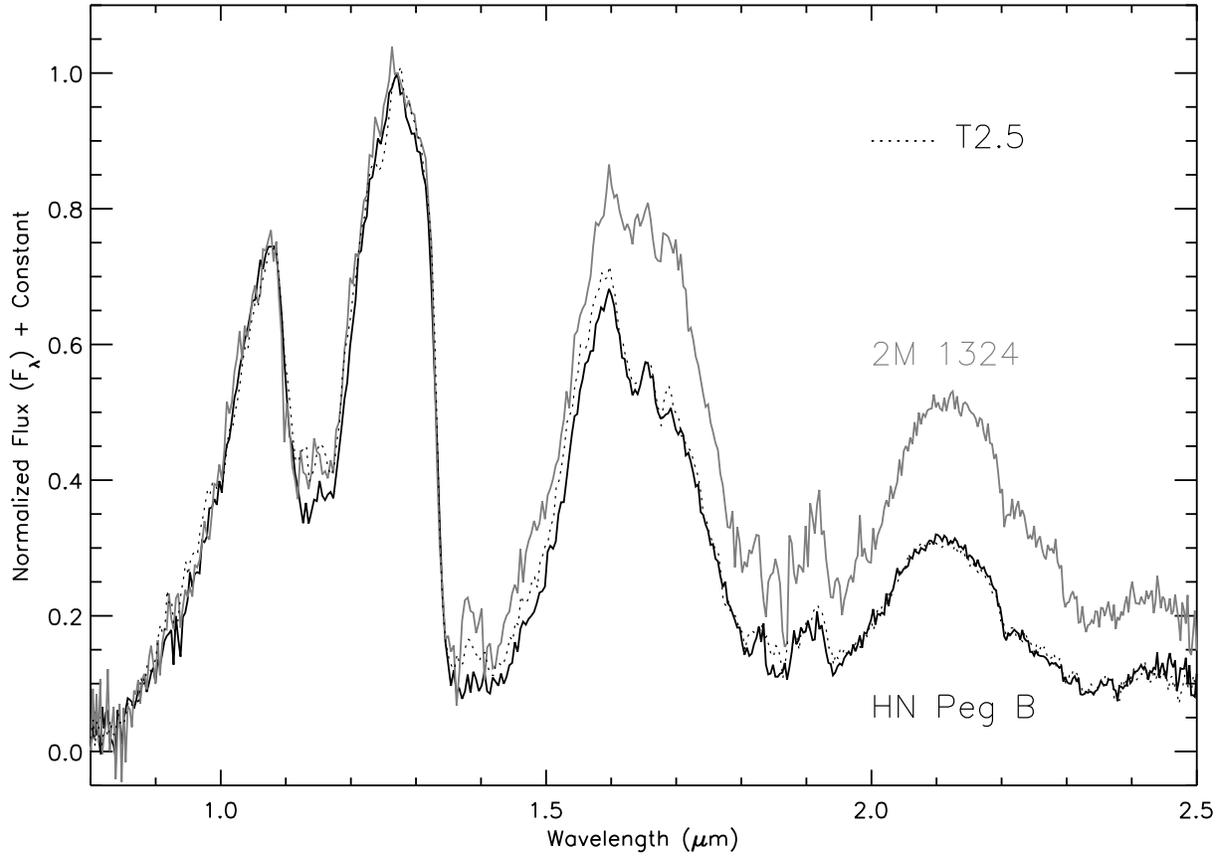}
\caption{2MASS 1324+63 (T2.5:pec, solid gray line) overplotted in comparison 
to HN Peg B (T2.5, solid black line, \citealt{2007ApJ...654..570L}), 
an $\sim$300 Myr old T dwarf.  For
comparison, a spectrum of a T2.5 dwarf constructed from the T2 and T3
spectral standards is also overplotted (dotted line).  All spectra
have been normalized at 1.27 $\mu$m.
\label{fig8}}
\end{figure}

\begin{figure} 
\epsscale{0.6}
\plotone{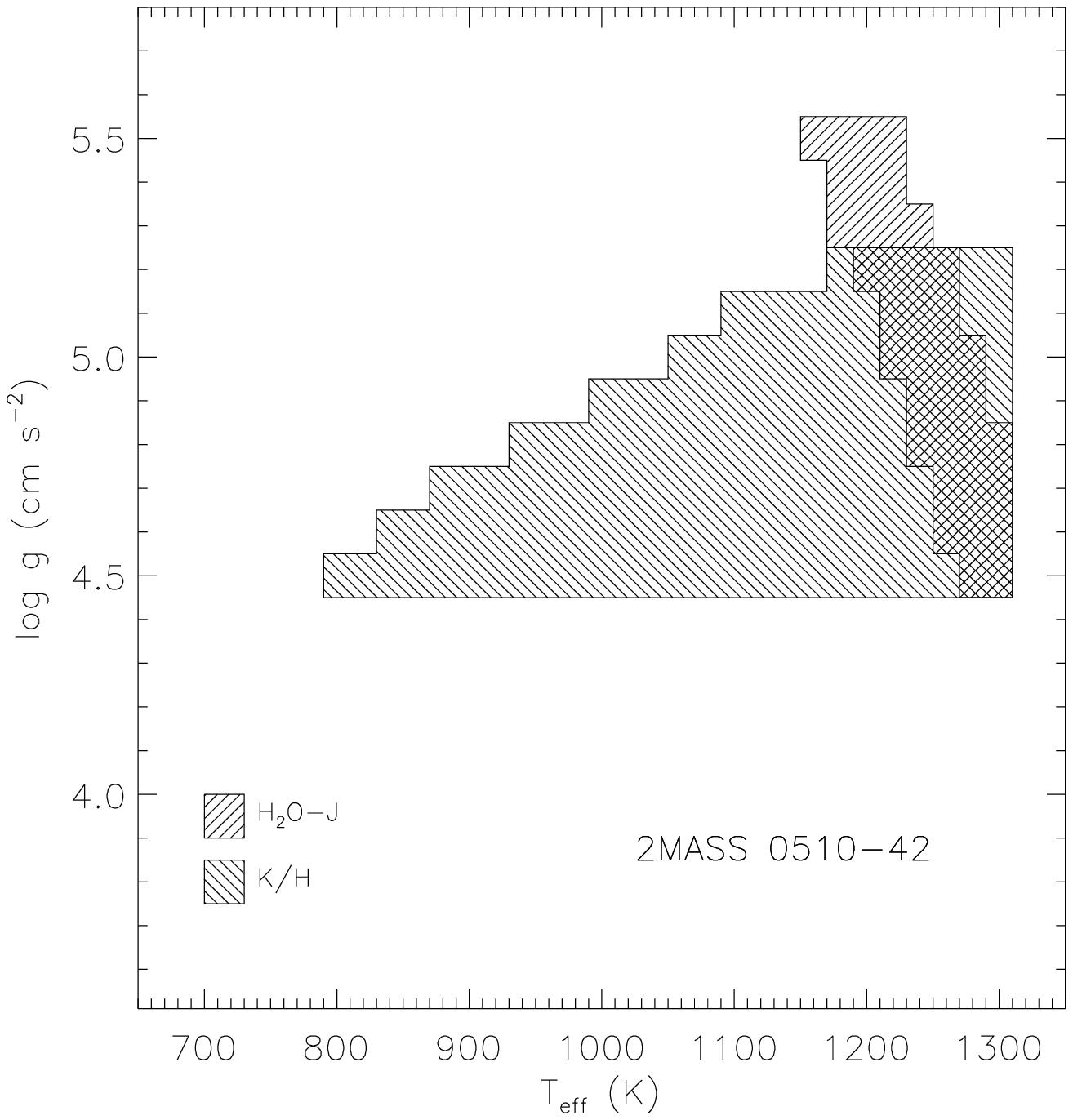}
\end{figure}

\begin{figure}
\epsscale{0.6}
\plotone{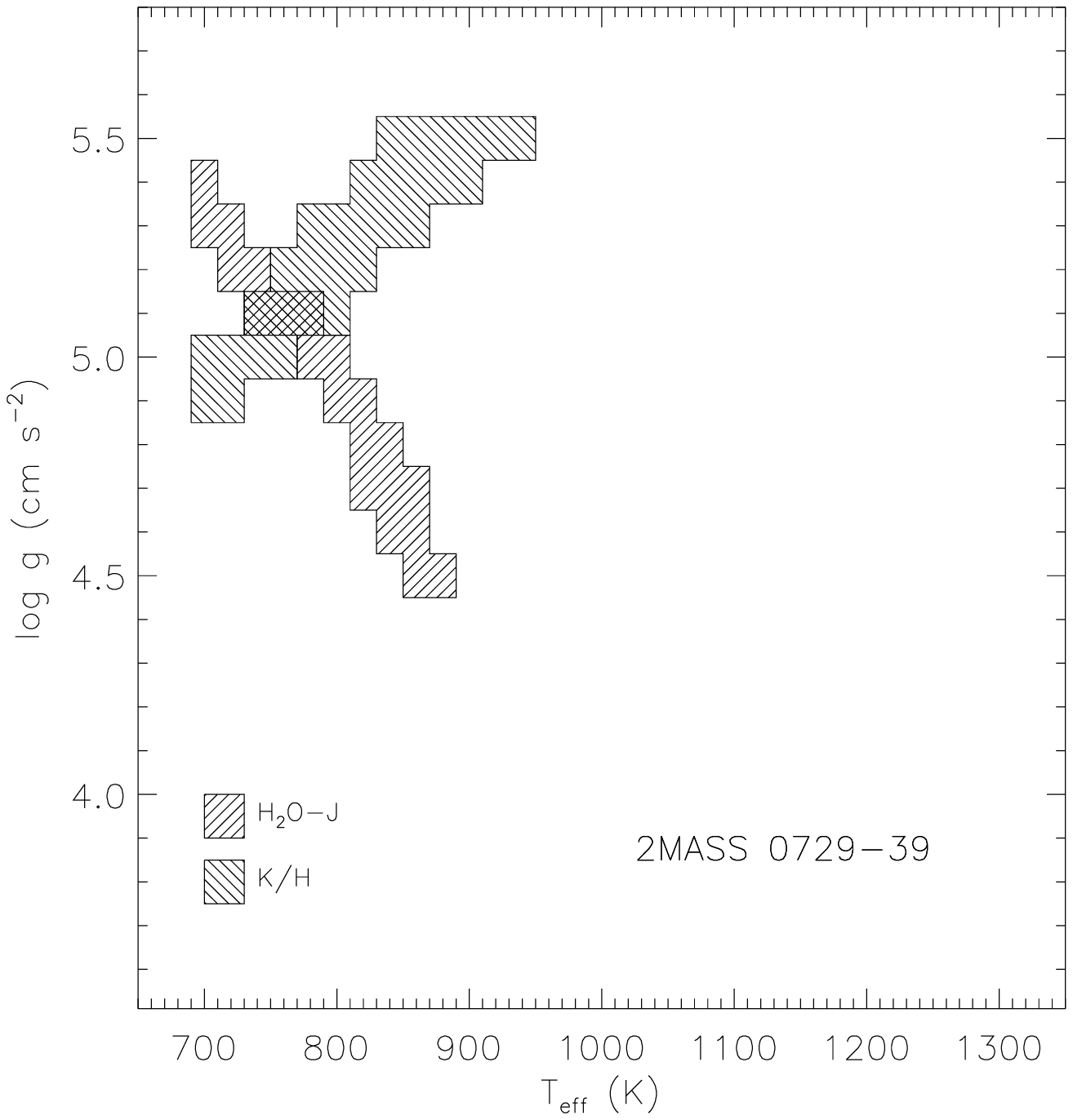}
\end{figure}

\begin{figure}
\epsscale{0.6}
\plotone{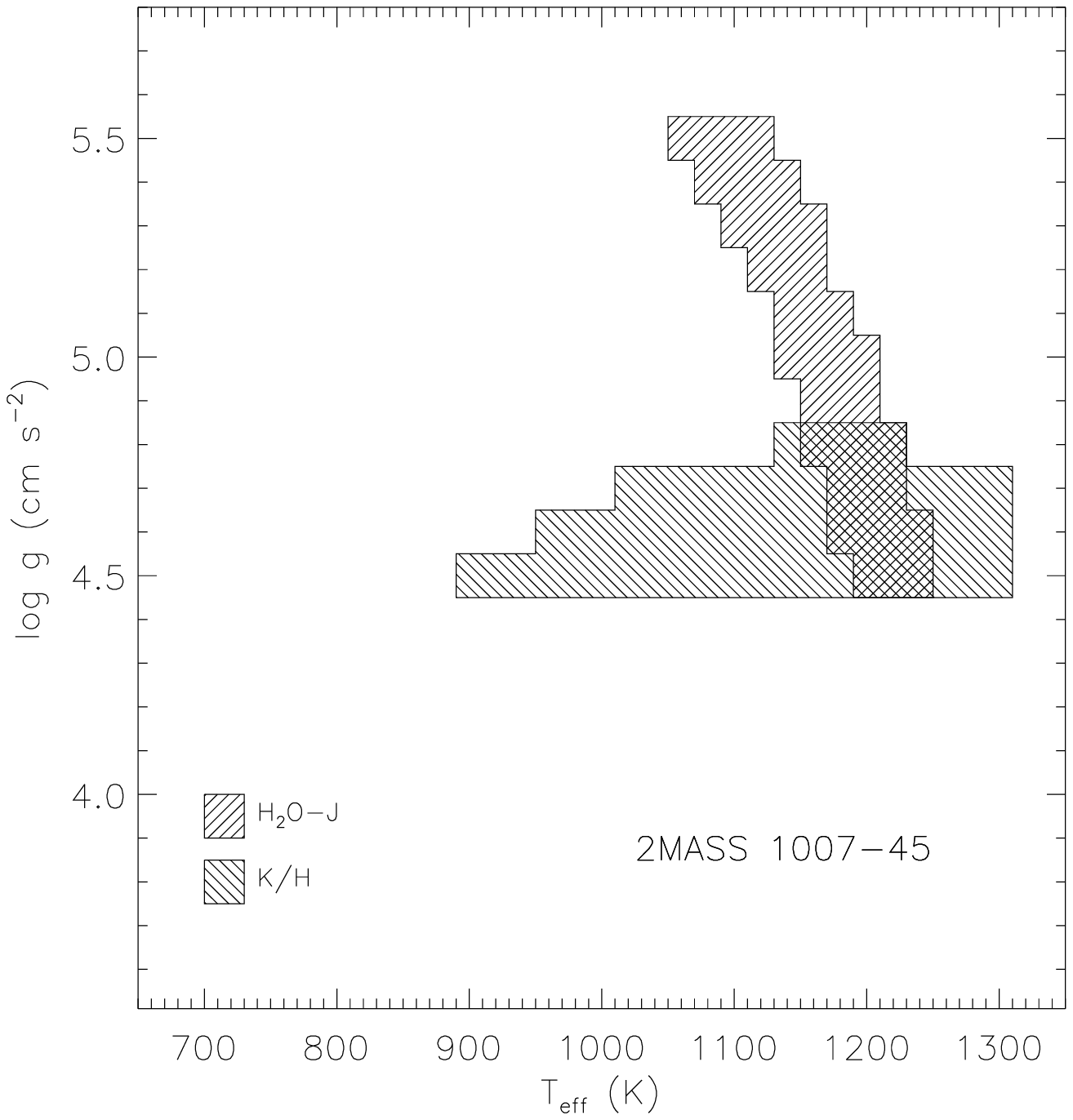}
\end{figure}

\begin{figure}
\plotone{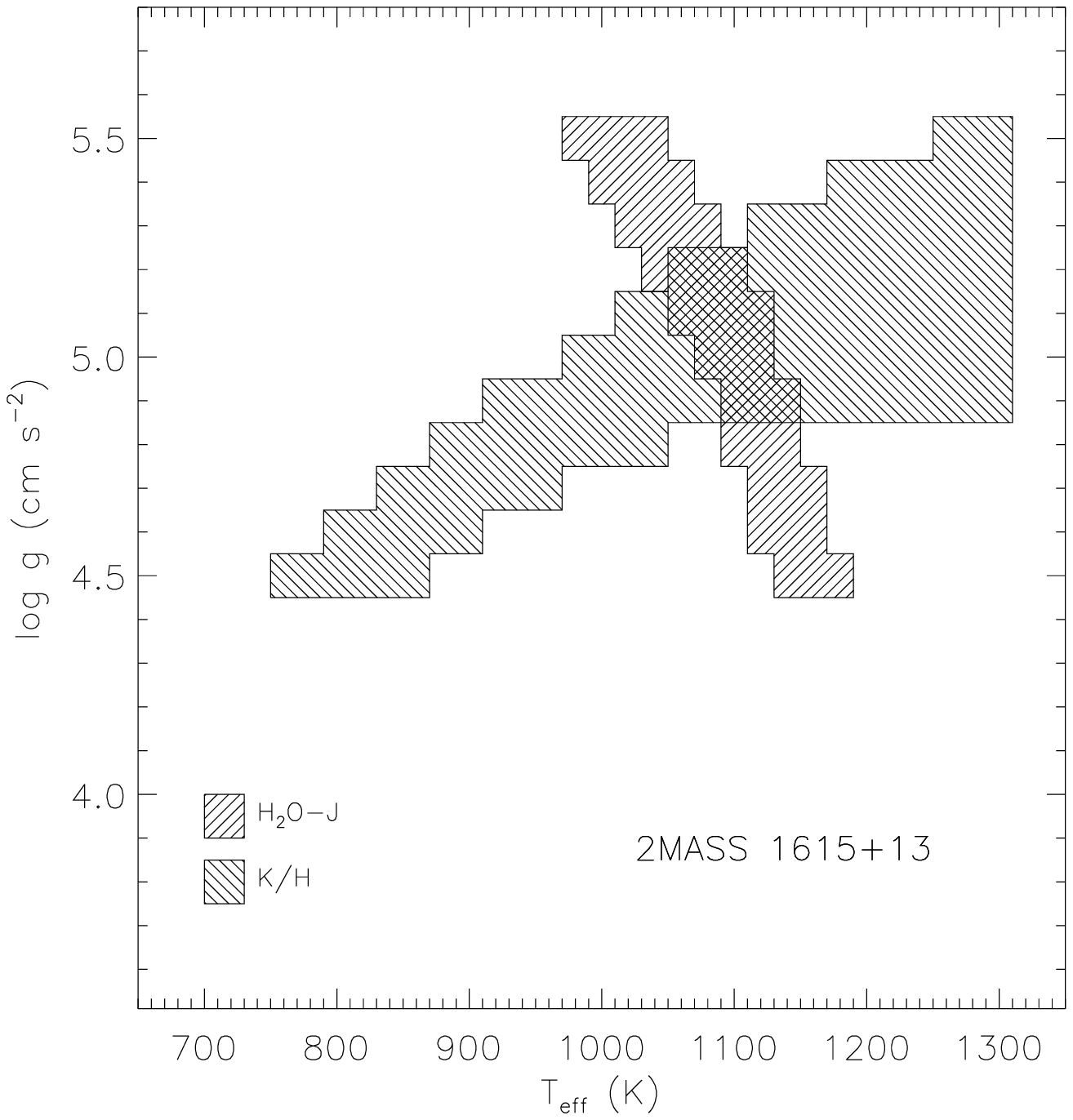}
\end{figure}

\begin{figure}
\plotone{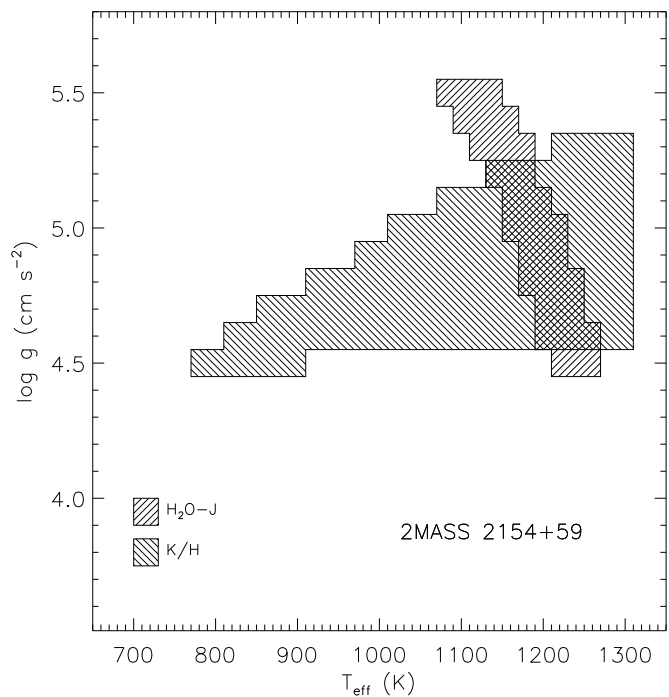}
\epsscale{0.6}
\caption{Spectral ratio measurements of H$_2$O and K/H (hatched regions) 
for 2MASS 0510$-$42 (T5), 2MASS 0729$-$39 (T8pec), 2MASS 1007$-$45 (T5), 
2MASS 1615+13 (T6), and 2MASS 2154+59 (T5) shown in T$_{eff}$ vs log g phase space.
Overlapping regions are the best fits; we list these values in Table 8.
\label{fig9}}
\end{figure}

\clearpage

\begin{deluxetable}{ccccccccc} 
\tabletypesize{\scriptsize}
\tablewidth{7.4in}
\tablenum{1}
\tablecaption{Description of 2MASS Photometric Selection of T Dwarfs}
\tablehead{ 
\colhead{} & \colhead{} & \colhead{} & \multicolumn{2}{c}{\# of Candidates} & 
\colhead{} & \colhead{} & \colhead{} & \colhead{} \\
\cline{4-5} \\
\colhead{Galactic Cut} & \colhead{J Mag Cut} & \colhead{Color Cuts} & 
\colhead{Cut 1\tablenotemark{a}} & \colhead{Cut 2\tablenotemark{b}} & 
\colhead{Followed-up} & \colhead{Transients} & \colhead{M Dwarfs} & 
\colhead{T Dwarfs}} 
\startdata
$|$b$|\ge$~15$^\circ$ & 16~$<$~J~$\le$~16.5\tablenotemark{c} & (J$-H\le$~0) or 
  & 13233 & 67[+3]\tablenotemark{d} & 56 & 44 & 8 & 4[+3] \\
& & (J$-H\le$~0.3, H$-K_s\le$~0) & & & & & & \\
10$^\circ$~$\le~|$b$|<$~15$^\circ$ & J~$\le$~16.5 & (J$-H\le$~0) or
  & 2719 & 8 & 6 & 4 & 1 & 1 \\
& & (J$-H\le$~0.3, H$-K_s\le$~0) & & & & & & \\
$|$b$|\le$~10$^\circ$, $|$l$|\ge$~20$^\circ$ & J~$\le$~16 & J$-H\le$~0 & 3711 & 6 & 5 & 2 & 0
  & 3 \\
\enddata
\tablenotetext{a}{Initial number of candidates from 2MASS Point Source Catalog.}
\tablenotetext{b}{Candidates remaining after visual inspection of DSS I
  \& II images.}
\tablenotetext{c}{J~=~16.5 has an SNR$\sim$10 in the 2MASS Point
Source Catalog, see Figure 7 in 
http://www.ipac.caltech.edu/2mass/releases/allsky/doc/sec2\_2.html\#pscphotprop.}
\tablenotetext{d}{Three previously known T dwarfs were identified.}
\end{deluxetable}

\begin{deluxetable}{lllllcl}
\tabletypesize{\scriptsize}
\tablewidth{5.2in}
\tablenum{2}
\tablecaption{Transients}
\tablehead{ 
\colhead{Object\tablenotemark{a}} & \colhead{J\tablenotemark{b}} & 
\colhead{H\tablenotemark{b}} & \colhead{K$_s$\tablenotemark{b}} &
\colhead{UT Date\tablenotemark{c}} & \colhead{$\beta$ (deg)\tablenotemark{d}}}
\startdata
2MASS J00082060$-$1922267 & 16.26$\pm$0.09 & 15.99$\pm$0.17 & 16.04(null)    & 1998 Aug 14 & $-$18.5 \\
2MASS J00511030$-$0704476 & 16.30$\pm$0.09 & 16.03$\pm$0.13 & 16.19$\pm$0.36 & 1998 Oct 02 & $-$11.6 \\
2MASS J01085000$-$0428510 & 16.11$\pm$0.08 & 16.23$\pm$0.19 & 15.92$\pm$0.27 & 1998 Sep 18 & $-$10.9 \\
2MASS J01153773+2445277   & 16.46$\pm$0.12 & 16.16$\pm$0.22 & 16.16(null)    & 1997 Oct 31 & +15.5   \\
2MASS J01202541$-$0421252 & 16.41$\pm$0.12 & 16.25$\pm$0.25 & 17.03(null)    & 1998 Sep 20 & $-$11.9 \\
2MASS J01354860+1309260   & 16.41$\pm$0.11 & 16.44$\pm$0.23 & 16.21(null)    & 2000 Sep 24 & +3.0    \\ 
2MASS J02113209+1959189   & 16.19$\pm$0.09 & 15.90$\pm$0.13 & 15.99(null)    & 1997 Oct 19 & +6.4    \\
2MASS J02200443+1829383   & 16.36$\pm$0.12 & 16.07$\pm$0.17 & 16.33$\pm$0.39 & 1997 Oct 20 & +4.3    \\
2MASS J02215590+1943340   & 16.21$\pm$0.10 & 16.26$\pm$0.22 & 15.64$\pm$0.22 & 1997 Oct 20 & +5.3    \\ 
2MASS J02242110+1926020   & 16.37$\pm$0.11 & 16.62$\pm$0.29 & 15.57(null)    & 1997 Oct 20 & +4.8    \\ 
2MASS J02262810+2327310   & 16.21$\pm$0.10 & 16.29$\pm$0.21 & 15.49(null)    & 1997 Oct 20 & +8.5    \\  
2MASS J02375362+2452399   & 16.19$\pm$0.09 & 15.93$\pm$0.15 & 16.01$\pm$0.25 & 1997 Nov 10 & +9.0    \\
2MASS J03532140$-$4154490 & 16.46$\pm$0.14 & 16.85$\pm$0.36 & 16.25(null)    & 1999 Sep 17 & $-$59.9 \\
2MASS J05314416$-$4224571 & 16.49$\pm$0.16 & 16.85$\pm$0.36 & 16.94(null)    & 1999 Oct 29 & $-$65.5 \\
2MASS J06550663+3024105   & 15.89$\pm$0.08 & 15.83$\pm$0.17 & 15.97$\pm$0.31 & 1998 Nov 23 & +7.5    \\
2MASS J08484742+0349184   & 16.36$\pm$0.15 & 16.40$\pm$0.28 & 16.91(null)    & 2000 Jan 29 & $-$13.5 \\
2MASS J09164433+0601113   & 16.31$\pm$0.13 & 16.24$\pm$0.24 & 16.83(null)    & 2000 Feb 22 & $-$9.3  \\
2MASS J10363196$-$1612231 & 16.27$\pm$0.11 & 16.03$\pm$0.15 & 16.34(null)    & 1998 Apr 02 & $-$23.1 \\
2MASS J11153870$-$0446599 & 16.42$\pm$0.10 & 16.92$\pm$0.30 & 16.95(null)    & 1999 Jan 04 & $-$8.8  \\
2MASS J11530357$-$2857301 & 16.42$\pm$0.15 & 16.74$\pm$0.36 & 16.75(null)    & 2000 Feb 01 & $-$27.0 \\
2MASS J12111689+1303508   & 16.31$\pm$0.11 & 16.05$\pm$0.18 & 16.52(null)    & 2000 Apr 05 & +13.1   \\
2MASS J12284242$-$1602593 & 16.44$\pm$0.12 & 16.28$\pm$0.19 & 16.80(null)    & 1998 Apr 01 & $-$11.9 \\
2MASS J12352384+2035252   & 16.14$\pm$0.10 & 15.89$\pm$0.18 & 15.89$\pm$0.24 & 1999 Apr 25 & +22.3   \\
2MASS J13175880$-$1427000 & 16.11$\pm$0.11 & 16.15$\pm$0.21 & 15.34(null)    & 1998 Apr 06 & $-$5.8  \\
2MASS J13271445$-$0819167 & 16.45$\pm$0.13 & 16.51$\pm$0.31 & 16.80(null)    & 1999 Feb 20 & +0.8    \\
2MASS J13343280$-$1952017 & 16.33$\pm$0.12 & 16.08$\pm$0.17 & 16.20(null)    & 1998 Apr 15 & $-$9.3  \\
2MASS J13502223$-$2210571 & 16.24$\pm$0.11 & 16.03$\pm$0.23 & 16.37(null)    & 1998 May 17 & $-$10.1 \\
2MASS J13540550+3139540   & 16.16$\pm$0.12 & 16.19$\pm$0.28 & 16.17(null)    & 1998 Mar 11 & +40.0   \\ 
2MASS J15452661$-$1418430 & 16.47$\pm$0.12 & 16.29$\pm$0.20 & 16.93(null)    & 1999 Mar 21 & +5.4    \\
2MASS J15470838$-$4111294 & 15.46$\pm$0.07 & 15.17$\pm$0.08 & 15.18$\pm$0.16 & 1999 May 14 & $-$20.7 \\
2MASS J16134510$-$3248360 & 15.56$\pm$0.07 & 15.63$\pm$0.14 & 15.36$\pm$0.21 & 1998 Jul 13 & $-$11.4 \\ 
2MASS J16490316+5357386   & 16.20$\pm$0.09 & 15.91$\pm$0.16 & 16.61(null)    & 1998 Jun 14 & +74.8   \\
2MASS J19404159+4752087   & 15.93$\pm$0.08 & 15.92$\pm$0.14 & 16.76(null)    & 1998 Jun 26 & +67.2   \\
2MASS J20043107$-$3710207 & 16.45$\pm$0.18 & 16.72$\pm$0.48 & 17.24(null)    & 1999 Jul 04 & $-$16.4 \\
2MASS J20320090$-$0749010 & 16.15$\pm$0.08 & 16.22$\pm$0.16 & 16.21(null)    & 1999 Jul 06 & +10.7   \\  
2MASS J20574228$-$0515055 & 16.20$\pm$0.10 & 15.92$\pm$0.21 & 16.12(null)    & 1998 Sep 20 & +11.5   \\
2MASS J21000824+4759033   & 15.79$\pm$0.08 & 15.93$\pm$0.15 & 14.72(null)    & 2000 Jun 10 & +60.4   \\
2MASS J21150144$-$4526057 & 16.34$\pm$0.11 & 16.36$\pm$0.26 & 17.23(null)    & 1999 Aug 13 & $-$28.0 \\
2MASS J21290545+4159005   & 15.60$\pm$0.06 & 16.09$\pm$0.20 & 15.29$\pm$0.16 & 2000 May 18 & +52.6   \\
2MASS J21542683+2116188   & 16.37$\pm$0.12 & 16.22$\pm$0.24 & 16.41(null)    & 1999 Oct 02 & +31.7   \\ 
2MASS J21573780$-$2047520 & 16.49$\pm$0.12 & 16.55$\pm$0.24 & 16.04(null)    & 1998 Aug 11 & $-$7.8  \\  
2MASS J22003181$-$2301592 & 16.43$\pm$0.14 & 16.34$\pm$0.28 & 16.96(null)    & 1998 Jul 04 & $-$10.2 \\
2MASS J22152050+1316290   & 16.46$\pm$0.13 & 16.25$\pm$0.19 & 16.47(null)    & 1997 Sep 13 & +22.4   \\
2MASS J22155422$-$2526573 & 16.14$\pm$0.11 & 15.85$\pm$0.14 & 16.08(null)    & 2000 Jul 24 & $-$13.7 \\
2MASS J22423443$-$0420011 & 16.39$\pm$0.10 & 16.18$\pm$0.18 & 16.46(null)    & 1998 Sep 29 & +3.6    \\
2MASS J22523538+2359146   & 16.20$\pm$0.11 & 15.91$\pm$0.20 & 15.94(null)    & 1997 Oct 06 & +28.6   \\
2MASS J23170461$-$0622477 & 16.13$\pm$0.09 & 15.96$\pm$0.19 & 16.15(null)    & 1998 Oct 10 & $-$1.6  \\
2MASS J23314348+1519291   & 16.39$\pm$0.12 & 16.22$\pm$0.20 & 16.32(null)    & 1997 Sep 29 & +16.8   \\
2MASS J23315967+1602183   & 16.29$\pm$0.12 & 16.26$\pm$0.21 & 16.85(null)    & 1997 Sep 29 & +17.5   \\
2MASS J23335590$-$0441137 & 16.04$\pm$0.07 & 15.77$\pm$0.12 & 15.87$\pm$0.23 & 2000 Sep 17 & $-$1.7  \\
\enddata 
\tablenotetext{a}{J2000 coordinates from the 2MASS All-Sky Point Source Catalog.}
\tablenotetext{b}{Photometry from 2MASS All-Sky Point Source Catalog.}
\tablenotetext{c}{UT date of 2MASS observation.}
\tablenotetext{d}{$\beta$: Ecliptic latitude.}
\end{deluxetable} 

\begin{deluxetable}{lllllcclcc}
\tabletypesize{\scriptsize}
\rotate
\tablewidth{8.8in}
\tablenum{3}
\tablecaption{IRTF/SpeX-prism Log}
\tablehead{ 
\colhead{Object\tablenotemark{a}} & \colhead{J\tablenotemark{b}} & 
\colhead{H\tablenotemark{b}} & \colhead{K$_s$\tablenotemark{b}} &
\colhead{UT Date} & \colhead{AM\tablenotemark{c}} & 
\colhead{N $\times$ t(s)\tablenotemark{d}} & \colhead{SpT} & 
\colhead{b$/\beta$ (deg)\tablenotemark{e}} & \colhead{Calibrator}}
\startdata
2MASS J01072340+4759060 & 15.90$\pm$0.08 & 16.11$\pm$0.18 &
  15.68$\pm$0.19 & 2006 Aug 17 & 1.15 & 8$\times$120 & M3 & $-$14.8$^\circ/$+37.2$^\circ$ & HD 9711 
  (A0 V) \\
2MASS J05063725$-$3405090 & 16.36$\pm$0.13 & 16.06$\pm$0.19 & 
  16.22(null) & 2006 Dec 08 & 1.70 & 8$\times$120 & 
  M7 & $-$35.5$^\circ/-$56.6$^\circ$ & HD 29028 (A0 V) \\
2MASS J05103520$-$4208140 & 16.22$\pm$0.09 & 16.24$\pm$0.16 & 
  16.00$\pm$0.28 & 2006 Sep 02 & 2.37 & 4$\times$120 & T5 & $-$35.9$^\circ/-$64.6$^\circ$ & HD 38921 
  (A0 V) \\
2MASS J05273570$-$1813490 & 16.49$\pm$0.23 & 16.53$\pm$0.39 & 
  15.16(null) & 2006 Sep 01 & 1.56 & 8$\times$120 & 
  M4 & $-$26.4$^\circ/-$41.4$^\circ$ & HD 37190 (A0 V) \\
2MASS J06020638+4043588 & 15.54$\pm$0.07 & 15.59$\pm$0.14 & 
  15.17$\pm$0.16 & 2006 Dec 09 & 1.07 & 8$\times$120 & T4.5 & +8.8$^\circ/$+17.3$^\circ$ & HD 39250 
  (A0 V) \\
2MASS J07290002$-$3954043 & 15.92$\pm$0.08 & 15.98$\pm$0.19 & 
  15.29(null) & 2006 Dec 08 & 2.00 & 16$\times$120 & T8.0pec & $-$10.4$^\circ/-$60.6$^\circ$ & HD 61516 
  (A0 V) \\
SDSSp J083717.22$-$000018.3 & 17.10$\pm$0.21 & 15.99$\pm$0.17 & 
  15.67(null) & 2006 Dec 21 & 1.08 & 20$\times$180 & 
  T1 std & +23.4$^\circ$/$-$17.9$^\circ$ & HD 74721 (A0 V) \\ 
2MASS J09162680+0628390 & 16.25$\pm$0.14 & 16.29$\pm$0.26 & 
  15.71(null) & 2006 Dec 09 & 1.03 & 4$\times$120 & M6.5 & +35.1$^\circ/-$8.9$^\circ$ & HD 79108 
  (A0 V) \\
2MASS J09212890$-$0339410 & 16.46$\pm$0.14 & 16.53$\pm$0.30 & 
  15.90(null) & 2006 Dec 09 & 1.12 & 8$\times$120 & 
  M7: & +30.8$^\circ/-$18.2$^\circ$ & HD 79108 (A0 V) \\
2MASS J09351549+0019103 & 16.47$\pm$0.15 & 16.27$\pm$0.20 & 
  16.61(null) & 2006 Dec 21 & 1.11 & 4$\times$120 & 
  M7: & +35.9$^\circ/-$13.3$^\circ$ & HD 74721 (A0 V) \\
2MASS J10073369$-$4555147 & 15.65$\pm$0.07 & 15.69$\pm$0.12 & 
  15.56$\pm$0.23 & 2006 Dec 09 & 2.45 & 8$\times$120 & T5 & +8.0$^\circ/-$52.1$^\circ$ & HD 88113 
  (A0 V) \\
2MASS J11061197+2754225 & 14.82$\pm$0.04 & 14.15$\pm$0.05 & 
  13.80$\pm$0.05 & 2006 Apr 08 & 1.03 & 8$\times$120 & T2.5 & +66.7$^\circ/$+20.3$^\circ$ & HD 89239 
  (A0 V) \\
SDSS J120747.17+024424.8 & 15.58$\pm$0.07 & 14.56$\pm$0.07 & 
  13.99$\pm$0.06 & 2006 Dec 20 & 1.21 & 8$\times$120 & T0 std & +63.5$^\circ$/+3.3$^\circ$ & HD 111744 
  (A0 V) \\
2MASS J12154432$-$3420591 & 16.24$\pm$0.13 & 15.81$\pm$0.19 & 
  16.32(null) & 2006 May 31 & 2.05 & 8$\times$120 & T4.5 & +27.9$^\circ/-$29.7$^\circ$ & SAO 203194 
  (A0 V) \\
2MASS J13243559+6358284 & 15.60$\pm$0.07 & 14.58$\pm$0.06 & 
  14.06$\pm$0.07 & 2006 Apr 11 & 1.49 & 6$\times$150 & T2.5 & +52.8$^\circ/$+62.6$^\circ$ & HD 118214 
  (A0 V) \\
2MASS J14044941$-$3159329 & 15.58$\pm$0.07 & 14.96$\pm$0.08 & 
  14.54$\pm$0.10 & 2006 Jun 01 & 2.18 & 12$\times$120 & T2.5 & +28.4$^\circ/-$18.1$^\circ$ & SAO 205525 
  (A0 V) \\ 
2MASS J16150413+1340079 & 16.35$\pm$0.09 & 16.49$\pm$0.25 & 
  15.86(null) & 2006 Sep 02 & 1.10 & 16$\times$120 & T6 & +48.8$^\circ/$+34.3$^\circ$ & q Her 
  (A0 V) \\
2MASS J16293700+1415440 & 16.27$\pm$0.11 & 16.36$\pm$0.26 & 
  15.00$\pm$0.15 & 2006 Sep 02 & 1.32 & 4$\times$120 & M6: & +37.8$^\circ/$+35.6$^\circ$ & q Her 
  (A0 V) \\
2MASS J20014670$-$3805400 & 16.42$\pm$0.11 & 16.44$\pm$0.23 & 
  15.87(null) & 2006 Sep 01 & 2.10 & 8$\times$120 & M4 & $-$29.5$^\circ/-$17.2$^\circ$ & HD 189501 
  (A0 V) \\
2MASS J21542494$-$1023022 & 16.43$\pm$0.12 & 16.45$\pm$0.28 & 
  17.05(null) & 2006 Aug 17 & 1.28 & 8$\times$120 & T4.5 & $-$45.2$^\circ/$+2.2$^\circ$ & HD 211278 
  (A0 V) \\
2MASS J21543318+5942187 & 15.66$\pm$0.07 & 15.77$\pm$0.17 & 
  15.34(null) & 2006 Nov 17 & 1.41 & 12$\times$180 & T5 & +4.1$^\circ/$+63.7$^\circ$ & BD+62 2148 
  (A0 V) \\
2MASS J22535460$-$0321450 & 16.01$\pm$0.09 & 16.09$\pm$0.22 & 
  15.46$\pm$0.23 & 2006 Aug 18 & 1.16 & 8$\times$120 & M6 & $-$53.2$^\circ/$+3.4$^\circ$ & HD 216807 
  (A0 V) \\ 
\enddata 
\tablenotetext{a}{J2000 coordinates from the 2MASS All-Sky Point Source Catalog.}
\tablenotetext{b}{Photometry from 2MASS All-Sky Point Source Catalog.}
\tablenotetext{c}{AM: Airmass.}
\tablenotetext{d}{Number of integrations times integration time.}
\tablenotetext{e}{b: Galactic latitude, $\beta$: Ecliptic latitude.}
\end{deluxetable} 

\begin{deluxetable}{llrcccccc} 
\tabletypesize{\scriptsize}
\rotate
\tablewidth{7.6in}
\tablenum{4}
\tablecaption{Spectrophotometric Properties of New and Confirmed T Dwarfs}
\tablehead{ 
\colhead{Object} & \colhead{J\tablenotemark{a}} & \colhead{J$-$H\tablenotemark{a}} & 
\colhead{H$-$K$_s$\tablenotemark{a}} & \colhead{J$-$K$_s$\tablenotemark{a}} & 
\colhead{i\tablenotemark{b}} & \colhead{z\tablenotemark{b}} &\colhead{SpT} & 
\colhead{Distance (pc)\tablenotemark{c}}} 
\startdata
\cutinhead{New T Dwarfs} 
2MASS J05103520$-$4208140 & 16.22$\pm$0.09 & $-$0.02$\pm$0.18 & 0.24$\pm$0.32 &
  0.23$\pm$0.30 & \nodata & \nodata & T5 & 24.3$\pm$2.0 \\
2MASS J06020638+4043588   & 15.54$\pm$0.07 & $-$0.05$\pm$0.15 & 0.43$\pm$0.21 &
  0.38$\pm$0.17 & \nodata & \nodata & T4.5 & 19.3$\pm$1.5 \\
2MASS J07290002$-$3954043 & 15.92$\pm$0.08 & $-$0.06$\pm$0.20 & $<$~0.69 &
  $<$~0.63 & \nodata & \nodata &T8pec & 8.4$\pm$0.7 \\
2MASS J10073369$-$4555147 & 15.65$\pm$0.07 & $-$0.04$\pm$0.14 & 0.13$\pm$0.26 &
  0.09$\pm$0.24 & \nodata & \nodata & T5 & 18.7$\pm$1.4 \\ 
2MASS J11061197+2754225   & 14.82$\pm$0.04 & 0.67$\pm$0.06 & 0.35$\pm$0.07 & 
  1.02$\pm$0.06 & 21.28$\pm$0.09 & 17.70$\pm$0.03 & T2.5 & 15.5$\pm$1.2 \\
2MASS J12154432$-$3420591 & 16.24$\pm$0.13 & 0.43$\pm$0.23 & $<-$0.51 &
  $<-$0.08 & \nodata & \nodata & T4.5 & 26.6$\pm$2.1 \\ 
2MASS J13243559+6358284   & 15.60$\pm$0.07 & 1.02$\pm$0.10 & 0.52$\pm$0.09 &
  1.54$\pm$0.10 & 22.68$\pm$0.26 & 18.73$\pm$0.04 & T2:pec & 21.8$\pm$1.7 \\
2MASS J14044941$-$3159329 & 15.58$\pm$0.07 & 0.62$\pm$0.10 & 0.42$\pm$0.12 &
  1.04$\pm$0.12 & \nodata & \nodata & T2.5 & 22.0$\pm$1.7 \\ 
2MASS J16150413+1340079   & 16.35$\pm$0.09 & $-$0.14$\pm$0.27 & $<$~0.63 & $<$~0.49 & 
  \nodata & 19.90$\pm$0.09 & T6 & 20.4$\pm$1.6 \\
2MASS J21542494$-$1023022 & 16.43$\pm$0.12 & $-$0.03$\pm$0.30 & $<-$0.59 &
  $<-$0.62 & \nodata & \nodata & T4.5 & 29.0$\pm$2.3 \\
2MASS J21543318+5942187   & 15.66$\pm$0.07 & $-$0.10$\pm$0.18 & $<$~0.43 &
  $<$~0.32 & \nodata & \nodata & T6 & 18.8$\pm$1.5 \\ 
\cutinhead{Recovered T Dwarfs} 
2MASS J09393548$-$2448279\tablenotemark{d} & 15.98$\pm$0.11 & 0.18$\pm$0.18 & 
  $<-$0.76 & $<-$0.58 & \nodata & \nodata & T8\tablenotemark{e} & 8.7$\pm$0.7 \\
SDSSp J134646.45$-$003150.4\tablenotemark{d} & 16.00$\pm$0.10 & 0.54$\pm$0.16 & 
  $-$0.31$\pm$0.30 & 0.23$\pm$0.29 & \nodata & \nodata & T6.5\tablenotemark{e} & 15.0$\pm$1.2 \\
SDSS J163022.92+081822.0\tablenotemark{d} & 16.40$\pm$0.11 & 0.07$\pm$0.31 & $<-$0.28 &
  $<-$0.21 & 23.82$\pm$0.61 & 20.13$\pm$0.10 & T5.5\tablenotemark{e} & 23.7$\pm$1.8 \\
SDSS J175805.46+463311.9\tablenotemark{d} & 16.15$\pm$0.09 & $-$0.10$\pm$0.24 & 0.79$\pm$0.29 & 
  0.69$\pm$0.21 & 24.18$\pm$0.57 & 19.67$\pm$0.07 & T6.5\tablenotemark{e} & 16.1$\pm$1.2 \\
SDSS J212413.89+010000.3\tablenotemark{d} & 16.03$\pm$0.07 & $-$0.15$\pm$0.21 & $<$~0.04 & 
  $<-$0.11 & 23.77$\pm$0.54 & 19.71$\pm$0.12 & T5\tablenotemark{e} & 22.3$\pm$1.7 \\
\enddata
\tablenotetext{a}{Photometry from 2MASS All-Sky Point Source Catalog.}
\tablenotetext{b}{Photometry from SDSS (AB) Data Release 5.}
\tablenotetext{c}{Spectrophotometric distance estimates derived in $\S$3.2.1.}
\tablenotetext{d}{References: 2MASS 0939$-$24 \citep{2005AJ....130.2326T}, SDSS 1346$-$00
  \citep{2000ApJ...531L..61T}, SDSS 1630+08 \citep{2006AJ....131.2722C}, SDSS 1758+46
  \citep{2004AJ....127.3553K}, SDSS 2124+01 \citep{2004AJ....127.3553K}.}
\tablenotetext{e}{\cite{2006ApJ...637.1067B}.}
\end{deluxetable}

\clearpage

\begin{deluxetable}{llllllllll} 
\tabletypesize{\scriptsize}
\tablewidth{7.3in}
\tablenum{5}
\tablecaption{Spectral Indices of 11 New T Dwarfs\tablenotemark{a}}
\tablehead{ 
\colhead{} & \colhead{} & \colhead{} & \colhead{} & \colhead{} & \colhead{} & 
\colhead{} & \colhead{} & \multicolumn{2}{c}{Derived NIR SpT} \\
\cline{9-10} \\
\colhead{Object} & \colhead{H$_2$O-{\it J}} & \colhead{CH$_4$-{\it J}} & 
\colhead{H$_2$O-{\it H}} & \colhead{CH$_4$-{\it H}} & 
\colhead{CH$_4$-{\it K}} & \colhead{H2O-{\it K}} & \colhead{{\it K/J}} &
\colhead{Direct\tablenotemark{b,c}} & \colhead{Ind 1\tablenotemark{c,d,e}}} 
\startdata
2MASS J1324+63 & 0.431 (T2.5) & 0.529 (T3) & 0.485 (T2) & 1.006 (L9.5) & 0.665 (T1.5) & 0.589 & 0.501 & T2:pec & T1.5: \\
2MASS J1106+27 & 0.481 (T2) & 0.564 (T2.5) & 0.522 (T1.5) & 0.781 (T2.5) & 0.614 (T2) & 0.585 & 0.336 & T2.5 & T2 \\
2MASS J1404$-$31 & 0.442 (T2.5) & 0.522 (T3) & 0.506 (T1.5) & 0.749 (T3) & 0.649 (T1.5) & 0.613 & 0.312 & T2.5 & T2.5 \\ 
2MASS J2154$-$10 & 0.349 (T4) & 0.438 (T4.5) & 0.395 (T4) & 0.524 (T4.5) & 0.273 (T4.5) & 0.498 & 0.212 & T4.5 & T4.5 \\
2MASS J0602+40 & 0.303 (T4.5) & 0.433 (T4.5) & 0.363 (T4.5) & 0.457 (T4.5) & 0.240 (T4.5) & 0.469 & 0.181 & T4.5 & T4.5 \\
2MASS J1215$-$34 & 0.308 (T4.5) & 0.486 (T4) & 0.382 (T4) & 0.414 (T5) & 0.151 (T6) & 0.536 & 0.225 & T4.5 & T4.5 \\ 
2MASS J0510$-$42 & 0.261 (T5) & 0.414 (T4.5) & 0.337 (T5) & 0.348 (T5.5) & 0.125 (T6.5) & 0.459 & 0.189 & T5 & T5.5 \\
2MASS J2154+59 & 0.217 (T5.5) & 0.354 (T6) & 0.325 (T5.5) & 0.361 (T5.5) & 0.163 (T5.5) & 0.437 & 0.167 & T5 & T5.5 \\ 
2MASS J1007$-$45 & 0.205 (T5.5) & 0.360 (T5) & 0.310 (T5.5) & 0.314 (T6) & 0.151 (T6) & 0.402 & 0.199 & T5 & T5.5 \\ 
2MASS J1615+13 & 0.167 (T6) & 0.313 (T6.5) & 0.305 (T5.5) & 0.267 (T6.5) & 0.153 (T6) & 0.399 & 0.145 & T6 & T6 \\
2MASS J0729$-$39 & 0.053 (T7.5) & 0.251 (T7) & 0.175 ($>$T8) & 0.202 (T7) & 0.044 ($>$T8) & 0.325 & 0.091 & T8pec & T7 \\
\enddata 
\tablenotetext{a}{NIR indices are defined by \cite{2006ApJ...637.1067B}.}
\tablenotetext{b}{Direct spectral typing is done by visually comparing 
  objects against standards.}
\tablenotetext{c}{Colons mark estimates with standard deviations $\ge$~1
spectral type.}
\tablenotetext{d}{Index values outside the proscribed range (i.e., '$<$'
or '$>$') are not used to compute the average spectral type.}
\tablenotetext{e}{Spectral typing follows the convention of
  \cite{2006ApJ...637.1067B}, where Ind 1 (in
  parentheses) is computed by comparison of the indices to the indices of standards.}
\end{deluxetable} 

\clearpage

\begin{deluxetable}{llrcccc} 
\tabletypesize{\scriptsize}
\tablewidth{5.2in}
\tablenum{6}
\tablecaption{Kinematic Properties of 4 T Dwarfs with Multi-Epoch Measurements}
\tablehead{ 
\colhead{Object} & \colhead{SpT} & \colhead{$\mu$ ($\arcsec$/yr)} & 
\colhead{$\theta$ (deg)} & \colhead{$\Delta$t (yr)} & \colhead{Dist (pc)} & 
\colhead{V$_{tan}$\tablenotemark{a} (km s$^{-1}$)}} 
\startdata
2MASS 1106+27\tablenotemark{b} & T2.5   & 0.57$\pm$0.08 & 216.6$\pm$0.3 & 2.13 &
  15.5$\pm$1.2 & 42$\pm$2 \\
2MASS 1324+63\tablenotemark{b} & T2:pec & 0.43$\pm$0.08 & 232.8$\pm$0.3 & 1.04 &
  21.8$\pm$1.7 & 45$\pm$3 \\
2MASS 1404$-$31\tablenotemark{b} & T2.5   & 0.35$\pm$0.03 & 275.3$\pm$0.2 & 2.55 &
  22.0$\pm$1.7 & 36$\pm$2 \\
2MASS 1615+13\tablenotemark{c} & T6     & 0.48$\pm$0.05 & 139.0$\pm$0.2 & 6.22 & 
  29.0$\pm$2.3 & 66$\pm$4 \\ 
\enddata
\tablenotetext{a}{V$_{tan}$~=~4.74$\times$D$\times\mu^\prime$ and 
$\sigma_{V_{tan}}$~=~$\surd$(4.74~($\mu^\prime\times\sigma_D^2$ + 
D~$\times~\sigma_{\mu^\prime}^2$)), where $\mu^\prime$=$\mu\times$1 yr [$\arcsec$].}
\tablenotetext{b}{Both epochs are from the 2MASS Working Database.}
\tablenotetext{c}{The first epoch is from the 2MASS Point Source Catalog, and 
the second epoch is from the SDSS Catalog.}
\end{deluxetable}

\clearpage

\begin{deluxetable}{llllllllll} 
\tabletypesize{\scriptsize}
\tablewidth{6.9in}
\tablenum{7}
\tablecaption{Synthetic Spectral Indices\tablenotemark{a}}
\tablehead{ 
\colhead{} & \colhead{} & \colhead{} & \colhead{} & \colhead{} &
\colhead{} & \colhead{} & \colhead{} & \multicolumn{2}{c}{Derived NIR SpT} \\
\cline{9-10} \\
\colhead{SpT of} & \colhead{} & \colhead{} & 
\colhead{} & \colhead{} & \colhead{} & \colhead{} &
\colhead{} & \colhead{} \\
\colhead{A+B} & \colhead{H$_2$O-{\it J}} & \colhead{CH$_4$-{\it J}} & 
\colhead{H$_2$O-{\it H}} & \colhead{CH$_4$-{\it H}} & 
\colhead{CH$_4$-{\it K}} & \colhead{H2O-{\it K}} & \colhead{{\it K/J}} &
\colhead{Direct\tablenotemark{b,c}} & \colhead{Ind 1\tablenotemark{b,d,e}}} 
\startdata
\bf{L8 std} & \bf{0.706 (L8)} & \bf{0.735 (L8)} & \bf{0.705 (L8)} & \bf{1.077 (L8)} & \bf{0.881 (L8)} & 
  \bf{0.696} & \bf{0.743} & \bf{L8} & \bf{L8} \\
L8+L9 & 0.666 (L8.5) & 0.705 (L8.5) & 0.661 (L8.5) & 1.085 ($<$L8) & 0.887 ($<$L8) & 0.697 & 0.688 & L8.5 & L8.5 \\
L8+T0 & 0.657 (L8.5) & 0.685 (L9) & 0.655 (L8.5) & 1.012 (L9.5) & 0.840 (L9.5) & 0.689 & 0.570 & L9 & L9 \\
L8+T1 & 0.630 (L9) & 0.683 (L9) & 0.625 (L9) & 1.023 (L9.5) & 0.807 (T0) & 0.660 & 0.525 & L9.5 & L9.5 \\
L8+T2 & 0.550 (T1.5) & 0.631 (T1.5) & 0.566 (T1) & 0.981 (T1) & 0.732 (T1) & 0.624 & 0.490 & T0 & T1 \\
L8+T3 & 0.502 (T1.5) & 0.582 (T2) & 0.554 (T1) & 0.862 (T2.5) & 0.712 (T1) & 0.618 & 0.413 & T1 & T1.5 \\
L8+T4 & 0.478 (T2) & 0.580 (T2) & 0.527 (T1.5) & 0.798 (T2.5) & 0.652 (T1.5) & 0.611 & 0.404 & T1.5: & T2 \\
L8+T5 & 0.407 (T3) & 0.492 (T4) & 0.542 (T1) & 0.767 (T2.5) & 0.705 (T1) & 0.631 & 0.367 & T2: & T2.5: \\
L8+T6 & 0.404 (T3) & 0.526 (T3) & 0.554 (T1) & 0.802 (T2.5) & 0.750 (T0.5) & 0.644 & 0.418 & T2: & T2: \\
L8+T7 & 0.447 (T2.5) & 0.528 (T3) & 0.585 (T0.5) & 0.853 (T2.5) & 0.775 (T0) & 0.648 & 0.504 & T1: & T1.5: \\
L8+T8 & 0.522 (T1.5) & 0.583 (T2) & 0.620 (L9) & 0.919 (T2) & 0.833 (L9.5) & 0.672 & 0.577 & T0: & T1: \\
\hline
\bf{L9 std} & \bf{0.631 (L9)} & \bf{0.681 (L9)} & \bf{0.621 (L9)} & \bf{1.092 (L9)} & \bf{0.894 (L9)} & 
  \bf{0.698} & \bf{0.640} & \bf{L9} & \bf{L9} \\
L9+T0 & 0.626 (L9.5) & 0.664 (L9.5) & 0.616 (L9.5) & 1.022 (L9.5) & 0.847 (L9.5) & 0.690 & 0.533 & L9.5 & L9.5 \\
L9+T1 & 0.604 (T0.5) & 0.664 (L9.5) & 0.589 (T0.5) & 1.032 (L9.5) & 0.815 (T0) & 0.661 & 0.494 & L9.5 & T0 \\
L9+T2 & 0.531 (T1.5) & 0.618 (T1.5) & 0.535 (T1.5) & 0.990 (L9.5) & 0.740 (T1) & 0.625 & 0.464 & T0.5 & T1 \\
L9+T3 & 0.486 (T2) & 0.571 (T2) & 0.523 (T1.5) & 0.875 (T2) & 0.721 (T1) & 0.619 & 0.392 & T1.5 & T1.5 \\
L9+T4 & 0.463 (T2) & 0.567 (T2) & 0.495 (T2) & 0.815 (T2.5) & 0.662 (T1.5) & 0.613 & 0.383 & T2 & T2 \\
L9+T5 & 0.394 (T3.5) & 0.484 (T4) & 0.502 (T1.5) & 0.789 (T2.5) & 0.717 (T1) & 0.633 & 0.345 & T2.5: & T2.5: \\
L9+T6 & 0.387 (T3.5) & 0.514 (T3) & 0.506 (T1.5) & 0.826 (T2.5) & 0.762 (T0.5) & 0.646 & 0.387 & T2.5: & T2: \\
L9+T7 & 0.422 (T3) & 0.511 (T3.5) & 0.528 (T1.5) & 0.878 (T2) & 0.788 (T0) & 0.650 & 0.459 & T2: & T2: \\
L9+T8 & 0.485 (T2) & 0.558 (T2.5) & 0.555 (T1) & 0.942 (T1.5) & 0.846 (L9.5) & 0.675 & 0.515 & T1: & T1.5: \\
\hline
\bf{T0 std} & \bf{0.621 (T0)} & \bf{0.651 (T0)} & \bf{0.612 (T0)} & \bf{ 0.955 (T0)} & \bf{0.790 (T0)} & 
  \bf{0.680} & \bf{0.442} & \bf{T0} & \bf{T0} \\
T0+T1 & 0.602 (T0.5) & 0.651 (T0) & 0.585 (T0.5) & 0.969 (T0) & 0.755 (T0.5) & 0.648 & 0.414 & T0.5 & T0.5 \\
T0+T2 & 0.533 (T1.5) & 0.610 (T1.5) & 0.533 (T1.5) & 0.933 (T1.5) & 0.677 (T1.5) & 0.610 & 0.396 & T1 & T1.5\\
T0+T3 & 0.491 (T2) & 0.566 (T2.5) & 0.521 (T1.5) & 0.820 (T2.5) & 0.647 (T1.5) & 0.601 & 0.332 & T2 & T2 \\
T0+T4 & 0.469 (T2) & 0.562 (T2.5) & 0.493 (T2) & 0.756 (T3) & 0.574 (T2) & 0.592 & 0.320 & T2.5: & T2.5 \\
T0+T5 & 0.405 (T3) & 0.483 (T4) & 0.499 (T1.5) & 0.718 (T3) & 0.615 (T2) & 0.610 & 0.279 & T3 & T2.5 \\
T0+T6 & 0.402 (T3) & 0.511 (T3.5) & 0.504 (T1.5) & 0.742 (T3) & 0.656 (T1.5) & 0.623 & 0.303 & T3: & T2.5 \\
T0+T7 & 0.436 (T2.5) & 0.508 (T4) & 0.524 (T1.5) & 0.779 (T2.5) & 0.679 (T1.5) & 0.626 & 0.347 & T2: & T2.5: \\
T0+T8 & 0.495 (T2) & 0.548 (T2.5) & 0.549 (T1) & 0.831 (T2.5) & 0.739 (T1) & 0.653 & 0.376 & T1: & T2 \\
\hline
\bf{T1 std} & \bf{0.584 (T1)} & \bf{0.653 (T1)} & \bf{0.563 (T1)} & \bf{0.981 (T1)} & \bf{0.724 (T1)} & \bf{0.620} & 
  \bf{0.391} & \bf{T1} & \bf{T1} \\
T1+T2 & 0.523 (T1.5) & 0.613 (T1.5) & 0.515 (T1.5) & 0.947 (T1.5) & 0.650 (T1.5) & 0.584 & 0.377 & T1.5 & T1.5 \\
T1+T3 & 0.485 (T2) & 0.573 (T2) & 0.504 (T1.5) & 0.840 (T2.5) & 0.616 (T2) & 0.572 & 0.318 & T2 & T2 \\
T1+T4 & 0.464 (T2) & 0.570 (T2) & 0.476 (T2) & 0.782 (T2.5) & 0.545 (T2.5) & 0.561 & 0.307 & T2.5 & T2 \\
T1+T5 & 0.406 (T3) & 0.499 (T4) & 0.478 (T2) & 0.752 (T3) & 0.576 (T2) & 0.571 & 0.267 & T3 & T3 \\
T1+T6 & 0.402 (T3) & 0.526 (T3) & 0.478 (T2) & 0.778 (T2.5) & 0.610 (T2) & 0.578 & 0.287 & T3 & T2.5 \\
T1+T7 & 0.433 (T2.5) & 0.528 (T3) & 0.494 (T2) & 0.817 (T2.5) & 0.629 (T1.5) & 0.578 & 0.323 & T2.5 & T2.5 \\
T1+T8 & 0.483 (T2) & 0.565 (T2.5) & 0.513 (T1.5) & 0.867 (T2.5) & 0.680 (T1.5) & 0.599 & 0.343 & T2 & T2 \\
\hline
\bf{T2 std} & \bf{0.474 (T2)} & \bf{0.583 (T2)} & \bf{ 0.474 (T2)} & \bf{0.917 (T2)} & \bf{0.585(T2)} & 
  \bf{0.552} & \bf{0.365} & \bf{T2} & \bf{T2} \\
T2+T3 & 0.442 (T2.5) & 0.547 (T2.5) & 0.463 (T2.5) & 0.818 (T2.5) & 0.546 (T2.5) & 0.537 & 0.312 & T2.5 & T2.5 \\
T2+T4 & 0.422 (T3) & 0.544 (T2.5) & 0.435 (T3.5) & 0.763 (T3) & 0.474 (T3) & 0.524 & 0.302 & T3 & T3 \\
T2+T5 & 0.365 (T4) & 0.477 (T4) & 0.429 (T3.5) & 0.732 (T3) & 0.486 (T3) & 0.525 & 0.266 & T3 & T3.5 \\
T2+T6 & 0.356 (T4) & 0.498 (T4) & 0.423 (T3.5) & 0.754 (T3) & 0.507 (T3) & 0.527 & 0.283 & T3 & T3.5 \\
T2+T7 & 0.373 (T4) & 0.494 (T4) & 0.429 (T3.5) & 0.785 (T2.5) & 0.518 (T3) & 0.524 & 0.313 & T3 & T3.5 \\
T2+T8 & 0.407 (T3) & 0.521 (T3) & 0.441 (T3) & 0.825 (T2.5) & 0.554 (T2.5) & 0.537 & 0.329 & T2.5 & T3 \\
\hline
\bf{T3 std} & \bf{0.413 (T3)} & \bf{ 0.516 (T3)} & \bf{0.453 (T3)} & \bf{0.717 (T3)} & \bf{0.496 (T3)} & \bf{0.519} & 
  \bf{0.264} & \bf{T3} & \bf{T3} \\
T3+T4 & 0.392 (T3.5) & 0.510 (T3.5) & 0.423 (T3.5) & 0.654 (T3.5) & 0.408 (T3.5) & 0.503 & 0.252 & T3.5 & T3.5 \\
T3+T5 & 0.337 (T4) & 0.445 (T4.5) & 0.414 (T3.5) & 0.602 (T4) & 0.405 (T3.5) & 0.497 & 0.215 & T4 & T4 \\
T3+T6 & 0.323 (T4.5) & 0.459 (T4.5) & 0.406 (T3.5) & 0.606 (T4) & 0.420 (T3.5) & 0.497 & 0.222 & T4 & T4 \\
T3+T7 & 0.334 (T4.5) & 0.449 (T4.5) & 0.411 (T3.5) & 0.619 (T3.5) & 0.427 (T3.5) & 0.490 & 0.240 & T4 & T4 \\
T3+T8 & 0.359 (T4) & 0.468 (T4.5) & 0.422 (T3.5) & 0.647 (T3.5) & 0.464 (T3) & 0.503 & 0.245 & T3.5 & T3.5 \\
\hline
\bf{T4 std} & \bf{0.369 (T4)} & \bf{0.506 (T4)} & \bf{0.389 (T4)} & \bf{0.581 (T4)} & \bf{0.305 (T4)} & \bf{0.482} & 
  \bf{0.239} & \bf{T4} & \bf{T4} \\
T4+T5 & 0.310 (T4.5) & 0.437 (T4.5) & 0.372 (T4.5) & 0.508 (T4.5) & 0.269 (T4.5) & 0.471 & 0.199 & T4.5 & T4.5 \\
T4+T6 & 0.291 (T4.5) & 0.450 (T4.5) & 0.356 (T4.5) & 0.498 (T4.5) & 0.266 (T4.5) & 0.466 & 0.204 & T4.5 & T4.5 \\
T4+T7 & 0.297 (T4.5) & 0.438 (T4.5) & 0.355 (T5)   & 0.499 (T4.5) & 0.260 (T4.5) & 0.455 & 0.220 & T4.5 & T4.5 \\
T4+T8 & 0.319 (T4.5) & 0.456 (T4.5) & 0.362 (T4.5) & 0.518 (T4.5) & 0.283 (T4) & 0.467 & 0.223 & T4 & T4.5 \\
\hline
\bf{T5 std} & \bf{0.240 (T5)} & \bf{0.356 (T5)} & \bf{0.345 (T5)} & \bf{0.393 (T5)} & \bf{0.200 (T5)} & 
  \bf{0.450} & \bf{0.151} & \bf{T5} & \bf{T5} \\
T5+T6 & 0.205 (T5.5) & 0.355 (T5.5) & 0.319 (T5.5) & 0.357 (T5.5) & 0.180 (T5.5) & 0.438 & 0.148 & T5.5 & T5.5 \\
T5+T7 & 0.195 (T5.5) & 0.322 (T6.5) & 0.311 (T5.5) & 0.332 (T5.5) & 0.159 (T6) & 0.417 & 0.155 & T5.5 & T6 \\
T5+T8 & 0.205 (T5.5) & 0.325 (T6.5) & 0.315 (T5.5) & 0.338 (T5.5) & 0.177 (T5.5) & 0.428 & 0.148 & T5.5 & T5.5 \\
\hline
\bf{T6 std} & \bf{0.154 (T6)} & \bf{0.354 (T6)} & \bf{0.280 (T6)} & \bf{0.301 (T6)} & \bf{0.149 (T6)} & \bf{0.418} & 
  \bf{0.142} & \bf{T6} & \bf{T6} \\
T6+T7 & 0.129 (T6.5) & 0.311 (T6.5) & 0.259 (T6.5) & 0.256 (T6.5) & 0.114 (T6.5) & 0.387 & 0.150 & T6.5 & T6.5 \\
T6+T8 & 0.127 (T6.5) & 0.313 (T6.5) & 0.255 (T6.5) & 0.250 (T6.5) & 0.127 (T6.5) & 0.394 & 0.140 & T6.5 & T6.5 \\
\hline
\bf{T7 std} & \bf{0.085 (T7)} & \bf{0.238 (T7)} & \bf{0.224 (T7)} & \bf{0.181 (T7)} & \bf{0.062 (T7)} & \bf{0.340} & 
  \bf{0.164} & \bf{T7} & \bf{T7} \\
T7+T8 & 0.070 (T7.5) & 0.219 (T7.5) & 0.209 (T7.5) & 0.153 (T7.5) & 0.059 (T7) & 0.332 & 0.152 & T7.5 & T7.5 \\
\hline
\bf{T8 std} & \bf{0.041 (T8)} & \bf{0.182 (T8)} & \bf{0.183 (T8)} & \bf{0.104 (T8)} & \bf{0.050 (T8)} & \bf{0.311} & 
  \bf{0.131} & \bf{T8} & \bf{T8} \\
\enddata 
\tablenotetext{a}{NIR indices are defined by \cite{2006ApJ...637.1067B}.}
\tablenotetext{b}{Colons mark estimates with standard deviations $\ge$~1
spectral type.}
\tablenotetext{c}{Direct spectral typing is done by visually comparing 
  objects against standards.}
\tablenotetext{d}{Index values outside the proscribed range (i.e., '$<$'
or '$>$') are not used to compute the average spectral type.}
\tablenotetext{e}{Spectral typing follows the convention of
  \cite{2006ApJ...637.1067B}, where Ind 1 (in
  parentheses) is computed by comparison of the indices to the indices of standards.}
\end{deluxetable}

\begin{deluxetable}{llcccc}
\tabletypesize{\scriptsize}
\tablewidth{4.5in}
\tablenum{8}
\tablecaption{T$_{eff}$ and Log g Estimates for Late-Type T Dwarfs}
\tablehead{
\colhead{Object} & \colhead{SpT} & \colhead{T$_{eff}$ [K]} &
\colhead{log g [cgs]} & \colhead{Mass [M$_\odot$]} & \colhead{Age [Gyr]}}
\startdata
2MASS 0510$-$42 & T5 & 1200--1300 & 4.5--5.2 & 0.017--0.049 & 0.13--2.1 \\
2MASS 0729$-$39 & T8pec & 740--780 & 5.1 & 0.038--0.039 & 3.8--4.1 \\
2MASS 1007$-$45 & T5 & 1160--1240 & 4.5--4.8 & 0.017-0.027 & 0.2--0.4 \\
2MASS 1615+13 & T6 & 1060--1140 & 4.9--5.2 & 0.030--0.048 & 0.6--2.9 \\
2MASS 2154+59 & T5 & 1140--1240 & 4.7--5.2 & 0.023--0.048 & 0.3--2.5 \\
\enddata
\end{deluxetable}

\clearpage

\begin{deluxetable}{lcccccccccc}
\tabletypesize{\scriptsize}
\tablewidth{8.5in}
\tablenum{9}
\rotate
\tablecaption{T Dwarfs in the Solar Neighborhood (within 25 pc)}
\tablehead{ 
\colhead{} & \colhead{} & \colhead{} & \colhead{} & \colhead{} &
\multicolumn{3}{c}{Distance (pc)} & \colhead{} & \colhead{} &
\colhead{} \\
\cline{6-8} \\
\colhead{Name of} & \colhead{Discovery} & \colhead{NIR} & 
\colhead{$J$\tablenotemark{a}} & \colhead{$K_s$\tablenotemark{a}} & 
\colhead{from} & \colhead{from} & \colhead{Adopted} & 
\colhead{High-res} & \colhead{Follow-up} & \colhead{} \\
\colhead{T Dwarf} & \colhead{Reference\tablenotemark{c}} & \colhead{SpT} & 
\colhead{(mag)} & \colhead{(mag)} & \colhead{$J$} & 
\colhead{$\pi_{trig}$} & \colhead{Distance\tablenotemark{b}} &
\colhead{imaging?} & \colhead{Reference\tablenotemark{c}} & \colhead{Remarks}} 
\startdata
$\epsilon$ Indi Bab (2MASS J2204$-$5646) & 1 & T1.0/T6.0 &
  12.29 & 11.35 & 4.45 & 3.63  & 3.6 & Yes & 28 & binary: a=$0{\farcs}732$ \\
SCR J1845$-$6357B & 2 & T5.5 & 13.16\tablenotemark{d} & \nodata & \nodata
  & 3.85 & 3.9 & Yes & 2 & single \\
2MASSI J0415195$-$093506 & 3 & T8.0 & 15.70 & 15.43
  & 7.58 & 5.74 & 5.7 & Yes & 29 & single \\
Gl 229B (J0610$-$2152) & 4 & T7.0p & 14.20 & 14.30 &
  5.53 & 5.77 & 5.8 & Yes & 4,30 & single \\
Gl 570D (2MASS J1457$-$2121) & 5 & T7.5 & 15.32 &
  15.24 &  7.73 & 5.91 & 5.9 & Yes & 31 & single \\
2MASSI J0937347+293142 & 3 & T6.0p & 14.65 & 15.27
  & 9.32 & 6.14 & 6.1 & Yes & 31 & single \\
IPMS J013656.57+093347.3 & 6 & T2.5 & 13.46 & 12.56 &
  8.27 &  \nodata & 8.3 & No & \nodata & single \\
\bf{2MASS J07290002$-$3954043} & \bf{7} & \bf{T8.0p} &
  \bf{15.92} & \nodata & \bf{8.41} & \nodata & \bf{8.4} & 
  \bf{No} & \nodata & \nodata \\
2MASS J15031961+2525196 & 8 & T5.0 & 13.94 & 13.96
  & 8.49 & \nodata & 8.5 & Yes & 29 & single \\
2MASS J09393548$-$2448279 & 9 & T8.0 & 15.98 & \nodata
  & 8.65 & \nodata & 8.7 & No & \nodata & \nodata \\
2MASSI J0727182+171001 & 3 & T7.0 & 15.60 & 15.56 &
  10.53 & 9.08 & 9.1 & No & \nodata & \nodata \\
2MASS J03480772$-$6022270 & 10 & T7.0 & 15.32 &
  15.60 &  9.25 & \nodata & 9.3 & Yes & 29 & single \\
2MASS J11145133$-$2618235 & 9 & T7.5 & 15.86 & \nodata & 9.89 &  
  \nodata & 9.9 & No & \nodata & \nodata \\
\hline
2MASS J05591914$-$1404488 & 11 & T4.5 & 13.80 &
  13.58 & 8.66 & 10.24 & 10.2 & Yes & 31 & single \\
2MASS J12373919+6526148 & 12 & T6.5 & 16.05 & \nodata
  & 15.33 & 10.41 & 10.4 & Yes & 31 & single \\ 
2MASSI J1047538+212423 & 12 & T6.5 & 15.82 & \nodata
  & 13.76 & 10.56 & 10.6 & Yes & 31 & single \\ 
2MASSI J0243137$-$245329 & 3 & T6.0 & 15.38 &
  15.22 & 13.06 & 10.68 & 10.7 & Yes & 29 & single \\ 
SDSSp J162414.37+002915.6 & 13 & T6.0 & 15.49 & \nodata &
  13.76 & 11.00 & 11.0 & Yes & 29 & single \\ 
2MASSI J1217110$-$031113 & 12 & T7.5 & 15.86 &
  \nodata & 9.90 & 11.01 & 11.0 & Yes & 29,27 & single \\ 
HD 3651B (J0039+2115) & 14 & T8 & 16.16\tablenotemark{d} & 16.87\tablenotemark{d} 
  & 9.39 & 11.11 & 11.1 & No & \nodata & \nodata \\ 
2MASSI J1546291$-$332511 & 3 & T5.5 & 15.63 & 15.49 & 16.66 & 11.36  
  & 11.4 & Yes & 31 & single \\ 
2MASSI J1553022+153236AB & 3 & T7.0 & 15.83 & 15.51
  & 11.68 & \nodata & $>$11.7 & Yes & 29 & binary: a=$0{\farcs}349$ \\ 
SDSSp J125453.90$-$012247.4 & 15 & T2.0 & 14.89 & 13.84
  & 15.76 & 11.78 & 11.8 & Yes & 29 & single \\ 
2MASS J00345157+0523050 & 16 & T6.5 & 15.54 & \nodata & 12.08 & \nodata & 
  12.1 & No & \nodata & \nodata \\ 
2MASS J00501994$-$3322402 & 9 & T7.0 & 15.93 & 15.24 & 12.24 & 
  \nodata & 12.2 & No & \nodata & \nodata \\ 
2MASS J12255432$-$2739466AB & 12 & T6.0 & 15.26 & 15.07 & 12.35 & 13.32  & 13.3 & Yes 
  & 31 & binary: a=$0{\farcs}282$ \\ 
2MASS J18283572$-$4849046 & 16 & T5.5 & 15.18 & 15.18 & 13.50 & 
  \nodata & 13.5 & No & \nodata & \nodata \\ 
2MASSI J1534498$-$295227AB & 3 & T5.5 & 14.90 & 14.84 & 11.90 & 13.59  
  & 13.6 & Yes & 31 & binary: a=$0{\farcs}65$ \\ 
2MASSI J2356547$-$155310 & 3 & T5.5 & 15.82 &
  15.77 & 18.21 & 14.50  & 14.5 & Yes & 31 & single \\ 
SDSSp J134646.45$-$003150.4 & 17 & T6.5 & 16.00 & 15.77 & 14.96 & 
  14.64 & 14.6 & No & \nodata & \nodata \\ 
2MASS J22282889$-$4310262 & 10 & T6.0 & 15.66 &
  15.30 & 14.86 & \nodata & 14.9 & Yes & 29 & single \\ 
SDSSp J042348.57$-$041403.5AB & 18 & T0.0 & 14.47 & 12.93
  & 11.10 & 15.17  & 15.2 & Yes & 29 & binary: a=$0{\farcs}164$ \\ 
SDSS J000013.54+255418.6 & 19 & T4.5 & 15.06 & 14.84 & 15.48 & 
  \nodata & 15.5 & No & \nodata & \nodata \\ 
\bf{2MASS J11061197+2754225} & \bf{7} & \bf{T2.5} & \bf{14.82} & \bf{13.80} & \bf{15.51} 
  & \nodata & \bf{15.5} & \bf{No} & \nodata & \nodata \\ 
SDSS J162838.77+230821.1 & 20 & T7.0 & 16.46 & 15.87 & 15.63 & \nodata & 
  15.6 & No & \nodata & \nodata \\
SDSS J175805.46+463311.9 & 19 & T6.5 & 16.15 & 15.47 & 16.05 & \nodata & 
  16.1 & No & \nodata & \nodata \\ 
SDSS J075840.33+324723.4 & 19 & T2.0 & 14.95 & 13.88 & 16.17 & 
  \nodata & 16.2 & No & \nodata & \nodata \\ 
2MASS J12314753+0847331 & 16 & T5.5 & 15.57 & 15.22 & 
  16.20 & \nodata & 16.2 & No & \nodata & \nodata \\ 
2MASS J11220826$-$3512363 & 9 & T2.0 & 15.02 & 14.38 & 16.72 & 
  \nodata & 16.7 & No & \nodata & \nodata \\ 
2MASSI J2254188+312349 & 3 & T4.0 & 15.26 & 14.90 & 18.01 & 
  \nodata & 18.0 & Yes & 29 & single \\ 
2MASS J21392676+0220226 & 21 & T1.5 & 15.26 & 13.58 & 
  18.18 & \nodata & 18.2 & No & \nodata & \nodata \\ 
SDSS J152039.82+354619.8 & 20 & T0.0 & 15.54 & 14.00 & 18.21 & 
  \nodata & 18.2 & No & \nodata & \nodata \\ 
HN Peg B (J2144+1446) & 22 & T2.5 & 15.86\tablenotemark{e} & 
  15.12\tablenotemark{e} & 25.03 & 18.39 & 18.4 \\ 
SDSS J120747.17+024424.8 & 23 & T0.0 & 15.58 & 13.99 & 18.54 & 
  \nodata & 18.5 & No & \nodata & \nodata \\
\bf{2MASS J10073369$-$4555147} & \bf{7} & \bf{T5.0} & \bf{15.65} & \bf{15.56} & \bf{18.68} 
  & \nodata & \bf{18.7} & \bf{No} & \nodata & \nodata \\ 
2MASS J23312378$-$4718274 & 16 & T5.0 & 15.66 & 15.39 & 18.76 & \nodata & 
  18.8 & No & \nodata & \nodata \\
\bf{2MASS J21543318+5942187} & \bf{7} & \bf{T5.0} & \bf{15.66}
  & \nodata & \bf{18.78} & \nodata & \bf{18.8} & \bf{No} & \nodata & \nodata \\ 
2MASSW J0920122+351742AB & 24 & T0.0 & 15.63 &
  13.98 & 18.93 & \nodata & $>$18.9 & Yes & 32 & binary: a=$0{\farcs}07$ \\
\bf{2MASS J06020638+4043588} & \bf{7} & \bf{T4.5} & \bf{15.54}
  & \bf{15.12} & \bf{19.32} & \nodata & \bf{19.3} & \bf{No} & \nodata & \nodata \\ 
2MASSI J0755480+221218 & 3 & T5.0 & 15.728 & 15.75
  & 19.36 & \nodata & 19.4 & Yes & 29 & single \\ 
2MASS J05160945$-$0445499 & 10 & T5.5 & 15.98 &
  15.49 & 19.60 & \nodata & 19.6 & Yes & 29 & single \\ 
SDSS J150411.63+102718.4 & 20 & T7.0 & 17.03 & 17.02 & 20.36 & \nodata & 
  20.4 & No & \nodata & \nodata \\ 
\bf{2MASS J16150410+1340070} & \bf{7} & \bf{T6.0} & \bf{16.35}
  & \nodata & \bf{20.40} & \nodata & \bf{20.4} & \bf{No} & \nodata & \nodata \\
Gl 337CD (2MASS J0912+1459) & 25 & T0.0 & 15.51 & 14.04 & 17.97 & 20.48 
  & 20.5 & Yes & 33 & binary: a=$0{\farcs}53$ \\ 
2MASS J19010601+4718136 & 16 & T5.0 & 15.86 & 15.64 &  
  20.54 & \nodata & 20.5 & No & \nodata & \nodata \\ 
SDSS J015141.69+124429.6 & 18 & T1.0 & 16.57 & 15.18 &
  31.89 & 21.40  & 21.4 & Yes & 29 & single \\
\bf{2MASS J13243559+6358284} & \bf{7} & \bf{T2.0p} & \bf{15.60} & \bf{14.06} & \bf{21.80} 
  & \nodata & \bf{21.8} & \bf{No} & \nodata & \nodata \\
\bf{2MASS J14044941$-$3159329} & \bf{7} & \bf{T2.5} & \bf{15.58} & \bf{14.54} & \bf{21.97} 
  & \nodata & \bf{22.0} & \bf{No} & \nodata & \nodata \\
SDSS J032553.17+042540.1 & 20 & T5.5 & 16.25 & 16.53 & 22.20 & 
  \nodata & 22.2 & No & \nodata & \nodata \\ 
SDSS J212413.89+010000.3 & 19 & T5.0 & 16.03 & 16.14 & 22.26 & 
  \nodata & 22.3 & No & \nodata & \nodata \\ 
2MASS J21513839$-$4853542 & 26 & T4.0 & 15.73 & 15.43 & 22.34 & \nodata 
  & 22.3 & No & \nodata & \nodata \\ 
2MASS J04070885+1514565 & 16 & T5.0 & 16.06 & 15.92 & 22.51 & 
  \nodata & 22.5 & No & \nodata & \nodata \\ 
SDSS J151114.66+060742.9 & 20 & T0.0 & 16.02 & 14.54 & 22.67 & \nodata & 
  22.7 & No & \nodata & \nodata \\ 
SDSSp J092615.38+584720.9AB & 18 & T4.5 & 15.90 & 15.45 & 22.73 & \nodata & 
  $>$22.7 & Yes & 29 & binary: a=$0{\farcs}070$ \\ 
SDSS J105213.51+442255.7 & 20 & T0.5 & 15.96 & 14.57 & 23.08 & \nodata 
  & 23.1 & No & \nodata & \nodata \\ 
SDSSp J111010.01+011613.1 & 18 & T5.5 & 16.34 & 15.13 &
  23.12 & \nodata & 23.1 & Yes & 29 & single \\ 
SDSS J074149.15+235127.5 & 19 & T5.0 & 16.16 & 15.85 & 23.61 & \nodata & 
  23.6 & No & \nodata & \nodata \\ 
SDSS J163022.92+081822.0 & 20 & T5.5 & 16.40 & 16.61 & 23.69 & \nodata & 
  23.7 & No & \nodata & \nodata \\ 
SDSS J074201.41+205520.5 & 19 & T5.0 & 16.19 & 15.23 & 23.99 & 
  \nodata & 24.0 & No & \nodata & \nodata \\ 
\bf{2MASS J05103520$-$4208140} & \bf{7} & \bf{T5.0} & \bf{16.22} & \bf{16.00} & \bf{24.31} 
  & \nodata & \bf{24.3} & \bf{No} & \nodata & \nodata \\
2MASS J05185995$-$2828372AB & 27 & T1.0 & 15.98 & 14.16 & 24.32 & \nodata & $>$24.3 
  & Yes & 29 & binary: a=$0{\farcs}051$ \\ 
2MASSI J2339101+135230 & 3 & T5.0 & 16.24 & 16.15 & 24.50 & 
  \nodata & 24.5 & Yes & 29 & single \\ 
\enddata 
\tablenotetext{a}{Photometry from 2MASS All-Sky Point Source Catalog.}
\tablenotetext{b}{Distance adopted from trigonometric parallax if measured.}
\tablenotetext{c}{References: (1) \cite{2003A&A...398L..29S}, (2) \cite{2006ApJ...641L.141B}, 
(3) \cite{2002ApJ...564..421B}, (4) \cite{1995Natur.378..463N}, (5) \cite{2000ApJ...531L..57B}, 
(6) \cite{2006ApJ...651L..57A}, (7) this paper, (8) \cite{2003AJ....125..850B}, 
(9) \cite{2005AJ....130.2326T}, 
(10) \cite{2003AJ....126.2487B}, (11) \cite{2000AJ....120.1100B}, (12) \cite{1999ApJ...522L..65B}, 
(13) \cite{1999ApJ...522L..61S}, (14) \cite{2006MNRAS.373L..31M}, (15) \cite{2000ApJ...536L..35L}, 
(16) \cite{2004AJ....127.2856B}, (17) \cite{2000ApJ...531L..61T}, (18) \cite{2002ApJ...564..466G}, 
(19) \cite{2004AJ....127.3553K}, (20) \cite{2006AJ....131.2722C}, (21) Cruz et al.\, in prep,
(22) \cite{2007ApJ...654..570L}, (23) \cite{2002AJ....123.3409H}, (24) \cite{2000AJ....120..447K}, 
(25) \cite{2001AJ....122.1989W}, (26) \cite{2005AJ....130.2347E}, (27) \cite{2004ApJ...604L..61C}, 
(28) \cite{2004A&A...413.1029M}, (29) \cite{2006ApJS..166..585B}, 
(30) \cite{1998AJ....115.2579G}, (31) \cite{2003ApJ...586..512B}, 
(32) \cite{2001AJ....121..489R}, 
(33) \cite{2005AJ....129.2849B}.}
\tablenotetext{d}{H-band magnitude from \cite{2006ApJ...641L.141B}.}
\tablenotetext{e}{Photometry is from \cite{2007ApJ...654..570L}.}     
\end{deluxetable} 

\clearpage

 \end{document}